\documentclass[sigconf]{acmart}

\copyrightyear{2024}
\acmYear{2024}
\setcopyright{rightsretained}
\acmConference[ICS '24]{Proceedings of the 38th ACM International Conference on Supercomputing}{June 4--7, 2024}{Kyoto, Japan}
\acmBooktitle{Proceedings of the 38th ACM International Conference on Supercomputing (ICS '24), June 4--7, 2024, Kyoto, Japan}\acmDOI{10.1145/3650200.3656601}
\acmISBN{979-8-4007-0610-3/24/06}

\usepackage{amsfonts,amsthm} 
\usepackage{subcaption} 
\usepackage{listings}
\usepackage{xcolor}
\usepackage{xspace}
\usepackage{algorithm}
\usepackage{algpseudocode}
\usepackage{algcompatible}
\usepackage{multirow}
\usepackage{centernot}
\usepackage{enumitem} 
\usepackage{hyphenat} 

\def\BibTeX{{\rm B\kern-.05em{\sc i\kern-.025em b}\kern-.08em
    T\kern-.1667em\lower.7ex\hbox{E}\kern-.125emX}}

\algtext*{ENDIF}
\algtext*{ENDWHILE}

\newtheorem{theorem}{Theorem}

\newtheoremstyle{bfnote}%
  {}{}
  {\itshape}{}
  {\bfseries}{.}
  { }{\thmname{#1}\thmnumber{ #2}\thmnote{ (#3)}}
\theoremstyle{bfnote}
\newtheorem{defi}{Definition}

\algrenewcommand\algorithmiccomment[1]{\hfill \(\triangleright\) #1}
  
\AtBeginDocument{%
  \providecommand\BibTeX{{%
    \normalfont B\kern-0.5em{\scshape i\kern-0.25em b}\kern-0.8em\TeX}}}

\begin{document}

\author{Durga Mandarapu}
\affiliation{
    \institution{Purdue University}
    \city{West Lafayette}
    \state{IN}
    \country{USA}}
\email{dmandara@purdue.edu}

\author{Vani Nagarajan}
\affiliation{
    \institution{Purdue University}
    \city{West Lafayette}
    \state{IN}
    \country{USA}}
\email{nagara16@purdue.edu}

\author{Artem Pelenitsyn}
\affiliation{
    \institution{Purdue University}
    \city{West Lafayette}
    \state{IN}
    \country{USA}}
\email{apelenit@purdue.edu}

\author{Milind Kulkarni}
\affiliation{
    \institution{Purdue University}
    \city{West Lafayette}
    \state{IN}
    \country{USA}}
\email{milind@purdue.edu}

\renewcommand{\shortauthors}{Mandarapu, et al.}
\title{Arkade: k-Nearest Neighbor Search With Non-Euclidean Distances using GPU Ray Tracing}

\begin{abstract}
\balance
High-performance implementations of $k$-Nearest Neighbor Search ($k$NN) in low dimensions use tree-based data structures.
Tree algorithms are hard to parallelize on GPUs due to their irregularity. 
However, newer Nvidia GPUs offer hardware support for tree operations through ray-tracing cores. 
Recent works have proposed using RT cores to implement $k$NN search, but they all have a hardware-imposed constraint on the distance metric used in the search---the Euclidean distance. 
We propose and implement two reductions to support $k$NN for a broad range of distances other than the Euclidean distance: Arkade Filter-Refine and Arkade Monotone Transformation, each of which allows non-Euclidean distance-based nearest neighbor queries to be performed in terms of the Euclidean distance.
With our reductions, we observe that $k$NN search time speedups range between $1.6$x-$200$x and $1.3$x-$33.1$x over various state-of-the-art GPU shader core and RT core baselines, respectively.
In evaluation, we provide several insights on RT architectures' ability to efficiently build and traverse the tree by analyzing the $k$NN search time trends. 

\end{abstract}



\maketitle

\section{Introduction}

$k$-Nearest Neighbor Search ($k$NN) is the problem of finding points similar to a query point based on a desired distance function. 
Several commonly used distance functions include Euclidean distance ($L^2$ norm), Manhattan distance ($L^1$ norm), Chebyshev distance ($L^\infty$ norm), Minkowski distance ($L^p$ norm), and Cosine (or Angular) distance. 
$k$NN is used in diverse applications, including point cloud registration~\cite{qiu2009gpu}, facial recognition~\cite{vision, neural}, recommendation systems~\cite{adeniyi2016automated}, and more.

Na\"ively scanning the entire dataset for every $k$NN query is expensive, especially when the dataset contains millions of data points.
Due to the computational intensity and the wide applicability of $k$NN, many optimization techniques have been proposed in this space:
\emph{tree-based approaches}, such as kd-tree or ball tree~\cite{flann, tigris, bufferkdtree}; graphs, such as proximity graphs or $k$NN graphs~\cite{hnsw, song, nsg}; hashing, such as locality sensitive hashing~\cite{indyk1998lsh, pham2022falconn, huang2015query};  quantization, such as product quantization codes~\cite{faiss, scann, andre2016cache}.

Tree-based approaches to $k$NN work better and provide logarithmic guarantees in lower dimensions~\cite{friedman1977algorithm, bentley1975multidimensional}.
Low-dimensional data (two to three dimensions) is predominant in several applications, such as spatial query processing~\cite{pandey2018good} and astronomical data~\cite{parallax}, where tree-based approaches have gained popularity.
However, tree-based approaches can not be efficiently accelerated using GPUs, unlike the non-tree indexing methods.
Tree-based implementations on GPU run $k$NN queries in parallel by mapping each query to a GPU thread that traverses the tree.
These traversals are highly {\em irregular:} different traversals touch different parts of the tree, leading to control divergence, and the tree itself can be scattered around memory, leading to memory divergence~\cite{autoropes}.
Nevertheless, several recently proposed algorithmic approaches improve GPU efficiency for tree traversals, leading to fast nearest neighbor searches on low-dimensional data~\cite{autoropes, treelogy, tigris, nam2016parallel}. 

Modern GPUs do not just contain the shader cores used by prior approaches. 
They also have ray tracing (RT) cores that are built to accelerate ray tracing~\cite{nvidia, amd, intel}: identifying which objects in a scene are intersected by rays cast from a source such as the viewer's eye. 
Ray-tracing is an inherently irregular problem, and these ray tracing cores perform {\em hardware accelerated tree traversals}: they build a spatial tree called a {\em bounding volume hierarchy} over the objects in a scene, then each ray traverses that tree to find the objects it intersects.
While ray tracing is a highly specific algorithm, and it may seem that RT cores cannot be used to solve other problems, prior work has shown that by carefully constructing the objects in a scene and properly defining the rays, it is possible to find solutions to non-ray tracing problems by {\em reducing} them to ray tracing~\cite{evangelou, rtnn, trueknn, wald}. In particular, several prior papers have shown how to reduce $k$NN to ray tracing~\cite{zellman, rtnn, trueknn, evangelou} (see Subsection~\ref{sec:relatedrt},\ref{sec:$k$NN-on-rt}).

\begin{figure}[!ht]
  \centering
  \includegraphics[width=0.65\linewidth]{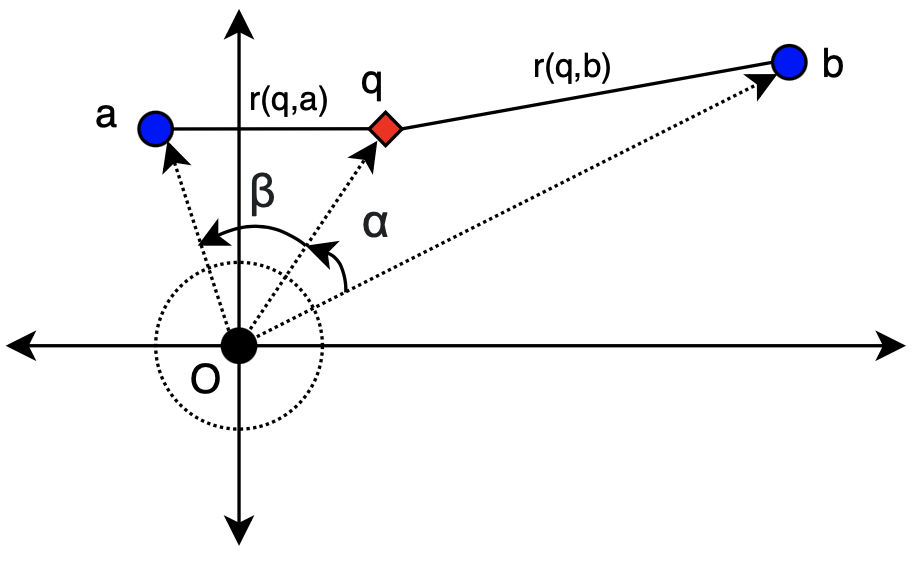}
  \caption{Euclidean and Angular distances: $a$ and $b$ are data points, $q$ is a query point, and $O$ is the point of reference. $L^2(q,a) < L^2(q,b) \centernot \implies \beta < \alpha$}
  \Description[Euclidean and Angular distances]{a and b are data points, q is a query point, and O is the point of reference. If the Euclidean distance between q and a is less than the Euclidean distance between q and b, then it does not imply angular distance between q and a is less than the angular distance between q and b.}
  \label{fig:l2-vs-cosine}
\end{figure}

Unfortunately, the existing $k$NN approaches on RT cores are all based around a single reduction that inherently uses the Euclidean distance ($L^2$ norm) as the desired distance function. 
This limitation is unsurprising, as the RT cores arrange objects in a scene according to the Euclidean distance.
However, one cannot merely use $L^2$ nearest neighbor as a proxy for other distance functions. 
For example, an object being at a particular Euclidean distance from the point of interest says nothing about a non-$L^2$ distance function such as the Angular distance between them (Figure~\ref{fig:l2-vs-cosine}). 
In practical applications like street maps and astronomical settings where Euclidean distance falls short in conveying essential information, non-$L^2$ distances are needed (see Section~\ref{sec:knn} for more detail).
However, prior work on RT cores cannot address these non-$L^2$ distance requirements.

To support non-Euclidean $k$NN queries, we make a key observation that the prior $k$NN reductions to RT cores do not solve the nearest neighbor problem \emph{directly}~\cite{evangelou, rtnn, trueknn}.
Instead, the reductions accelerate an {\em $r$-bounded distance query}: find all points within a distance $r$ of a query point $q$, according to their Euclidean distance. 
Similarly, instead of solving $k$NN problem for other distance functions on RT cores directly, in this paper, we show how to reduce $k$NN searches in other distances to the $r$-bounded Euclidean distance search and implement the reductions for RT cores.

\subsection*{Contributions}
This paper introduces {\em Arkade}\footnote{%
  The name, Arkade(aRKaDe), comes from the three parameters this paper considers -- radius ($r$), number of neighbors ($k$), and distance function ($D$).},
a suite of two general reductions: {\em Filter\hyp{}Refine} (FR) and {\em Monotone Transformation} (MT), each allowing non\hyp{}Euclidean distance\hyp{}based nearest neighbor queries to be performed on RT cores, despite the RT cores' inherent constraint of using Euclidean distances.

In particular, we contribute the following.

\begin{enumerate}[leftmargin=*]
\item \emph{Arkade FR reduction} performs a generic distance-based $k$NN search using RT cores by decoupling the $k$NN search into \textit{Filter} and \textit{Refine} phases and adapting a tree-based $k$NN algorithm for distances besides the Euclidean distance (Section~\ref{sec:filter-refine}).
The reduction utilizes geometric properties of the distances to exploit RT-core acceleration. 
We provide specifically optimized implementations for several different distance functions: $L^1$, $L^\infty$, and general $L^p$ distances.

\item \emph{Arkade MT reduction} enables RT-based acceleration of $k$NN search for distance functions that do not hold the geometric properties favored by Arkade FR reduction (Section~\ref{sec:design-monotone}). The reduction transforms the input such that the original order of distances between the data points is preserved. Important examples of such distances are cosine distance or angular distance.

\item Evaluation of Arkade (FR and MT) implemented as stand-alone applications using RT cores of the NVIDIA GeForce RTX 4060 Ti GPU (Section~\ref{sec:eval}). Our reductions show speedups of $1.6$x-$200$x and $1.3$x-$33.1$x over various state-of-the-art GPU shader core and RT core baselines, respectively. 

\end{enumerate}

\section{Background}

\subsection{k-Nearest Neighbour Search}
\label{sec:knn}
We define the $k$NN search problem in Definition~\ref{def:$k$NN} since there are several variants of $k$NN.
Importantly, the particular distance function $D$ is a parameter. 

\begin{defi} [$k$-Nearest Neighbor Search] Given a query point $q \in \mathbb{R}^d$, a set of data points, $A \subseteq \mathbb{R}^d$, a value $k \in \mathbb{N}$, and a distance function $D: \mathbb{R}^d \times \mathbb{R}^d \rightarrow \mathbb{R}$, the generalized $k$-nearest neighbor problem finds a result set of points, $T \subseteq A$, that contains the closest $k$ points to $q$ according to $D$.
\label{def:$k$NN}
\end{defi}

The naive way of performing $k$NN is computing the distance between $q$ and all of the points in $A$ and ordering them by $D$, an $O(n \log k)$ process\footnote{The $\log k$ term comes from efficiently maintaining distances of the top $k$ neighbors} ($|A| = n$).
Tree-based approaches~\cite{flann, tigris, bufferkdtree} can avoid comparing $q$ to every point in $A$ by efficiently indexing the points in $A$ using a tree and pruning the search space, resulting in an $O(\log n \log k)$ algorithm. 
However, the trees built by RT cores use $L^2$-based pruning~\cite{evangelou, rtnn, trueknn, wald}, and hence this approach {\em only} works if the distance function $D$ is the $L^2$ distance.

\subsubsection*{Other distance functions}

The key focus of this paper is using RT cores, which are inherently tied to $L^2$ distances, to solve non-$L^2$ distance problems. This subsection summarizes some of these non-$L^2$ distance functions.

In 2-dimensional space, we recall the set of functions that give the distance between point $a$ and $b$ based in $L^p$ spaces~\cite{bourbaki1987topological} as follows, where$a_x, a_y$ are $x,y$ coordinates of point $a$, and $|.|$ represents the absolute value:
\begin{align*}
    L^p(a,b) &= {(|a_x-b_x|^p+|a_y-b_y|^p)}^{\frac{1}{p}}, p \in \mathbb{R}\ge 1 \\
    L^\infty (a,b) &= max(|a_x-b_x|,|a_y-b_y|)   
\label{equation:lpnorm}
\end{align*}

While $L^p$ norms use Cartesian coordinates,
distances such as angular distance, cosine distance, inner product, or dot product use spherical coordinates~\cite{cosine}.
Angular distance is the shorter angle between two vectors, while cosine distance or cosine similarity is the cosine function applied to this angle.
The inner product or dot product is the same for vectors, and these are, in turn, the same as cosine distance when the vectors are of unit length.
Because cosine distance measures how similar two vectors are, it is highly useful in recommendation systems.

This paper confines its scope to 2- and 3-dimensional spaces because RT cores operate solely within these dimensions. 
The utility of non-Euclidean distances in these lower dimensions remains evident in several domains, such as
geospatial applications and astronomy. 
For instance, consider street maps, where the determination of nearest points of interest hinges on the ordering of their \textit{Manhattan distance} from the query point location since the data points in a city usually adhere to taxicab geometry.
Similarly, visually nearby stars are identified with cosine distance in 3 dimensions instead of Euclidean distance: the three stars in Orion's belt are not $L^2$-close together---they are approximately 2000, 1200, and 700 light years away from Earth---despite being visually adjacent.

\subsection{Ray Tracing Architecture} \label{sec:rt-cores}

Ray tracing is a graphics rendering algorithm where rays are modeled from a starting point as a source and followed (traced) till they hit the objects in a scene. 
The fundamental operation in ray tracing is computing {\em ray-object intersections}: for a given ray, what object(s) does the ray intersect? 
This problem shares some features with nearest-neighbor search: the na\"ive algorithm compares a ray to each object in a scene but can be accelerated using a spatial tree to prune the space. 
In the case of ray tracing, this spatial tree is called a {\em bounding volume hierarchy} (BVH)~\cite{bvh_first}, and the RT architecture on modern GPUs provides acceleration for building and traversing this spatial tree.

The RT architecture employs both RT cores and shader cores (also called streaming multiprocessors) to accelerate various stages of the ray-tracing pipeline. 
Optimized drivers build a BVH bottom-up by enclosing each object in an {\em axis-aligned bounding box} (AABB) and grouping AABBs such that several AABBs can be enclosed in a larger AABB.
Eventually, the overall scene is enclosed in a single AABB.
RT cores recursively traverse the BVH tree to compute ray-object intersections. 
In particular, if a ray intersects an AABB, then the enclosed bounding boxes will be tested next. 
The process continues until it reaches leaf AABBs. 
At that point, the shader cores execute user-defined code to determine whether the ray intersects the object contained in the AABB. 
If an intersection is found, another user-specified code is called.

\subsection{Programming and Execution Model} 
\label{sec:optix-prog-model}

Optix~\cite{optix} is a programming interface that provides access to the entire RT architecture.
This interface allows the user to write traditional shader programs that are executed on the shader cores and leverage the RT hardware for BVH construction, traversal, and, if applicable, intersection testing. 
Optix allows the user to specify user-defined geometries, which we use to represent neighborhoods in non-$L^2$ distances.
Important Optix kernels that we use are RayGen and Intersection. 
RayGen kernel creates rays with user-specified parameters such as the origin, direction, and length of the ray. 
It then calls for BVH traversal and intersection testing.
For user-defined geometries, the user is required to provide a custom intersection test for ray-object intersections in the form of an Intersection kernel.

\subsubsection{Geometric Objects}
For a distance function $D$, all the points that are at a $D$-distance of $r$ could be described by a geometric object.
For example, if the distance function is $L^1$ norm, then the geometric object is a square rhombus in 2D space and a square bi-pyramid in 3D space.
Similarly, if the distance is $L^2$ norm, then the geometric object is a circle in 2D space and a sphere in 3D space.
A geometric object simply refers to a geometry whose periphery contains points that are equidistant from the center of the geometry.
The geometric objects are then placed inside AABBs.
With the Optix interface, it is up to the user to define the distance function of geometric objects, so these geometric objects are also called user-defined or custom-defined geometries.

\subsubsection{Limitations}
There are several limitations when re-purposing RT architecture to perform a non-RT task.
First, we are limited to using data with three dimensions.
Second, the BVH built by the RT architecture is not accessible nor programmable in any kind by a user.
There is no available information on how the BVH is constructed or traversed.
Third, during the traversal, the Optix interface notifies the user only when a successful ray-AABB intersection occurs.
We do not know the actual number of AABBs that are tested during the traversal or the actual traversal path.
Fourth, even after successful mapping, it is hard to assess the resource utilization of our mapping and identify opportunities to optimize the hardware usage due to inadequate support from the profilers.

\subsection{RT-kNN: kNN on RT architecture}\label{sec:$k$NN-on-rt}

Accelerating a non-RT problem with RT cores requires defining several components that we call a \emph{reduction}.
A reduction defines a scene with objects and rays such that the hardware-accelerated ray-AABB intersection detection encodes a partial or complete solution to the initial non-RT problem. The reduction should define how to decode that solution.

\begin{figure}[!ht]
  \centering
  \includegraphics[width=0.9\linewidth]{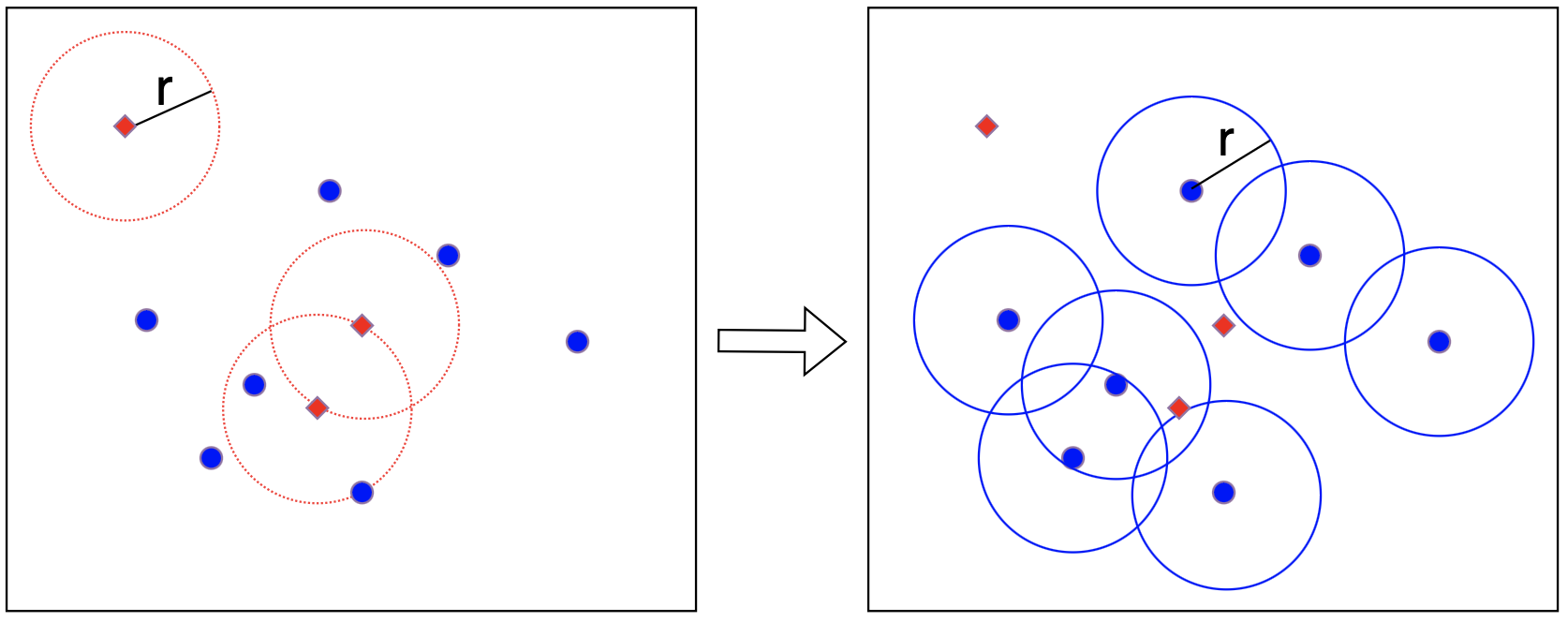}
  \caption{RT-$k$NN reduction (right) finds all query points within radius $r$ to data point unlike the conventional $k$NN algorithm (left) that finds all data points within radius $r$ to the query point. Blue circles and red rhombus represent data and query points, respectively.}
  \Description[compares rt-knn to conventional knn]{Image on the left shows a conventional way of performing $k$NN search which is to find all the data points within a radius r of the query point. The image on the right shows RT-$k$NN reduction which tries to find all query points within radius $r$ to data point.}
  \label{fig:rt-knn}
\end{figure}

In particular, the reduction of $k$NN to RT only aims to accelerate a part of the problem, which is the {\em $r$-bounded distance query}.
We refer to this reduction as `RT-$k$NN' for the rest of the paper.
Figure~\ref{fig:rt-knn} shows how the RT-$k$NN reduction (on the right) solves a flipped-around version of the conventional $k$NN algorithm (on the left).
It tries to find if the query point is at a distance less than or equal to $r$ to a data point rather than finding the data points that are within a distance of less than or equal to $r$ to a query point.
To find the neighbors of query points, RT-$k$NN reduction models the data points as spheres, the query points as rays, and the neighbor identification as an intersection of the corresponding ray with the spheres, as explained in more detail below~\cite{zellman}.

Given a set of data points $A$ and a set of query points $Q$, the RT-$k$NN reduction builds spheres of radius $r$ centered around all data points in $A$, as shown in the right part of Figure~\ref{fig:rt-knn}.
To find the neighbors, the reduction launches point rays from every query point.
A point ray is a ray whose length is a very small positive number.
If a point ray cast from the point $q\in Q$ as the source intersects with the sphere of radius $r$ and center $a\in A$, then it means that the point $q$ is present inside the sphere and so the Euclidean distance between points $a$ and $q$ is at most $r$.


\section{Filter-Refine} 
\label{sec:filter-refine}
In this section, we show how to map $k$-nearest neighbor search for distance functions beside $L^2$ norm to a ray tracing problem.
For this purpose, we use a general framework called Filter\hyp{}Refine. In particular, we formulate {\em Arkade Filter\hyp{}Refine reduction} (Subsec.~\ref{subsec:arkade-fr}) and prove its correctness (Subsec.~\ref{subsec:filter-refine-correctness}).

Filter\hyp{}Refine is a two-step selection framework for search problems~\cite{filter}.
First, we {\em filter} a subset of the possible candidates from the data points and then {\em refine} this subset to produce a result that answers the original search query, exactly or approximately.
Drawing on the principles of this framework, we devise a reduction that breaks down $k$NN search for a generic distance function into Filter and Refine phases and maps these phases to operations performed by the RT architecture.

\subsection{Arkade Filter\hyp{}Refine Reduction}%
\label{subsec:arkade-fr}

Assume an arbitrary distance function $D$.
To find the nearest $k$ points within the $D$-distance of $r$ to a query point $q$, the Arkade Filter\hyp{}Refine (FR) reduction employs the following {\em Filter} and {\em Refine} phases. 

\begin{enumerate}[leftmargin=*]
\item {\em Filter Phase} finds all the candidate data points that are within the $D$-distance of $r$, by mapping the data points and query points to a ray tracing scene. 
\item {\em Refine Phase}, once the candidates are filtered, sorts them according to their $D$-distance to the query point $q$ and finds the $k$ nearest neighbors.
\end{enumerate}

The first step of the reduction involves solving the $r$-bound query problem, which is where the RT architecture comes in.
The hardware accelerates the search process of candidates since we encode them as a part of the ray tracing problem.
In particular, to find all the data points within a $D$ distance of $r$ from the query points, the reduction builds specific distance function geometric objects centered at data points and launch point rays originating from query points. 
The ray traverses the BVH to find the candidates that will be passed to the Refine phase.

\begin{figure}[H]
  \centering
  \includegraphics[width=0.8\linewidth]{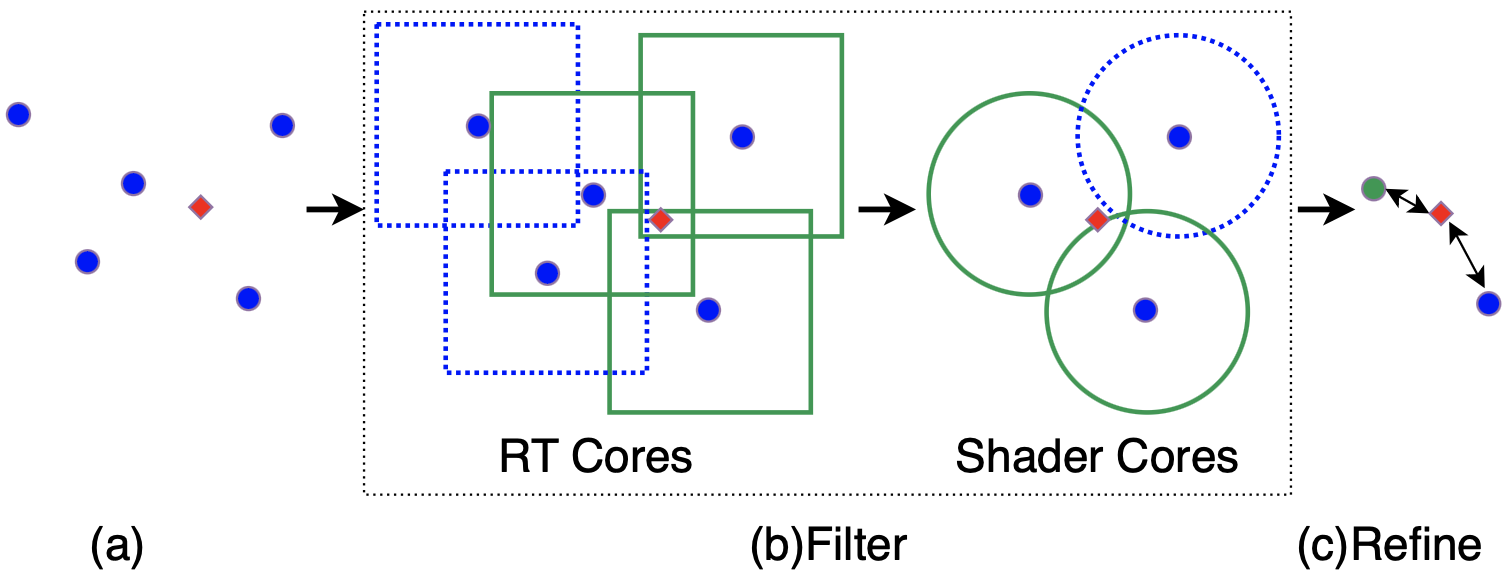}
  \vspace{-5pt}
  \caption{Filter\hyp{}Refine: (a) map points to RT scene, (b) RT cores filter AABBs and shader cores filter geometric objects, (c) refine candidates to select $k=1$ nearest neighbors.}
  \Description[shows filter-refine working]{first, data and query points are mapped to RT scene, second, AABBs are filtered by RT cores and geometric objects by shader cores, and third candidates are refined to select $k=1$ nearest neighbors.}
  \label{fig:fiter-refine}
\end{figure}

In Figure~\ref{fig:fiter-refine}, part (a) on the left shows data points and query points colored in blue and red, respectively.
When an RT core finds an intersection, the intersection is with the AABB that contains the geometric object, rather than the geometric object itself.
To ensure that there are no false positive candidates, the ray-AABB intersections are further filtered to remove the data points where the query point lies inside the AABB but outside the geometric object. 
Part (b) of Figure~\ref{fig:fiter-refine} shows how the Filter phase first uses RT cores to get the AABBs that a point ray intersects and then uses the shader cores to perform the intersection with the geometric object present inside these intersected AABBs.
AABBs and geometric objects are represented by squares and circles, respectively. 
The green AABBs or circles are the ones selected, while the blue ones are not.

The second step of the reduction, the Refine phase, processes the candidates that are passed on from the Filter step.
In particular, the candidates are ordered to select the nearest $k$ data points to the query point.
Part (c) in Figure~\ref{fig:fiter-refine} shows that the blue and green points are the candidates processed in the refine phase, out of which only the green points are selected as the top-$k$ neighbors.

We present {\em Arkade Filter\hyp{}Refine reduction} in algorithm~\ref{alg:arkade}. 
In Line~\ref{line:1}, AABBs corresponding to each data point are defined.
If $D$ is $L^1$ norm, then the geometric object is a square rhombus and the AABB with a side of length $2r$ should be defined to tightly fit the rhombus. 
It is up to the user to decide how big of a bounding box is needed to render the desired geometry. 
However, the tighter, the better.
In line~\ref{line:2}, an Optix API call is made to build the BVH on the defined AABBs.
The constructed BVH is not returned to the user but is available for the RT architecture to traverse.
In line~\ref{line:3}, an Optix API call is made to launch point rays from each query point.
From line~\ref{line:4}, the neighbor search starts.
In lines~\ref{line:4}-\ref{line:5}, RT cores perform the BVH tree traversal of the ray and return the AABB intersection when a ray is found to intersect with AABB.
Lines~\ref{line:5}-\ref{line:9} and lines~\ref{line:10}-\ref{line:12} indicate the Filter and Refine phases respectively.
Lines~\ref{line:6}-\ref{line:7} extract the data point, which is the center of geometry inside the hit AABB, and the query point, which is the source of the point ray hitting the AABB.
In line~\ref{line:8}, we compute the $D$-distance between them.
In line~\ref{line:9}, we filter out all the data points that are farther than a $D$-distance of $r$.
Lines~\ref{line:10}-\ref{line:12} refine the selected candidates and store the top $k$ closest points.

\begin{algorithm}[!ht]
\caption{Arkade Filter-Refine Reduction}\label{alg:arkade}
\textbf{Input:} Training set $A$, Query set $Q$, distance function $D$, $r, k$\\
\hspace{-0.5cm}\textbf{Output:} $\forall q \in Q,$ top $k$ neighbors of q within $D$ distance $r$ 
\begin{algorithmic}[1]
\STATE $\forall a\in A,$ define AABB on the geometry centered at $a$ \label{line:1}
\STATE construct BVH on all the AABBs \label{line:2}
\STATE $\forall q\in Q,$ launch point ray at $q$ \label{line:3}
\WHILE {each ray is  traversing BVH} \label{line:4}
  \IF {RT cores return ray-AABB intersection}    \label{line:5}
    \STATE $a \gets geometry.center$ \Comment{data point}\label{line:6}
    \STATE $q \gets ray.origin$ \Comment{query point} \label{line:7}  
    \STATE $w \gets D(a,q)$ \label{line:8} 
    \IF {$w \le r$}\label{line:9}
    \IF {$w < max(neighbors(q).distance)$ \label{line:10} \\ \hspace{1.2cm} or $|neighbors|< k$} \label{line:11}
        \STATE $neighbors.insert(a,w)$ \label{line:12}
    \ENDIF
    \ENDIF
  \ENDIF 
\ENDWHILE
\end{algorithmic}
\end{algorithm}

In the Algorithm~\ref{alg:arkade}, the Filter and Refine phases are interleaved.
Instead of storing all the candidates from the Filter phase, each candidate is refined on the go.
Once the RT core finds a point that is within $r$ distance, the reduction uses the shader cores to dynamically update the list of $k$ nearest neighbors, and return the control to RT cores to resume the search for candidates.

The Arkade FR reduction presented above is a generalization of the RT-$k$NN reduction and uses RT cores in a novel way.
RT-$k$NN ships spheres to RT cores because an $r, L^2$-ball (Def~\ref{def:r,d-ball}) is exactly a sphere.
Similarly, for a distance $D$, we need to build geometric objects customized to the distance function to represent an $r, D$-ball.
Our key observation with Arkade Filter\hyp{}Refine is that the RT architecture can process custom geometric objects.

Although Arkade FR reduction depends on a more advanced feature of RT cores, it finds a way to
stay agnostic to the inherent property of RT cores, which only understand $L^2$ distances.  
The distance fixed by the hardware does not impact the core idea of the reduction---using point rays to find the $k$NN candidates.
A point ray intersects with an object containing it as long as the query point is present inside the geometric object centered at a data point, and this does not depend on the hardware-defined metric. 

Arkade FR reduction is generic over the distance function $D$.
Hence, the effectiveness of this reduction depends on the distance function and consequently, the geometric objects that will be built centered at the data points.
If a distance function geometric object is such that the Filter phase forwards most of the data points as $k$NN candidates, Arkade FR reduction is not useful.
Because it has to process the unnecessarily large number of candidates in the Refine phase and this might not be better than a linear scan.
An example of such a distance function is cosine distance.
We address how to perform cosine distance-based $k$NN search in Section~\ref{sec:design-monotone}.

\subsection{Correctness of Arkade FR Reduction}%
\label{subsec:filter-refine-correctness}

To prove the correctness of Arkade FR reduction, we first introduce $r, D$-ball in Definition~\ref{def:r,d-ball} and then formally define Filter and Refine phases in Definitions~\ref{def:filter} and \ref{def:refine} respectively.

\begin{defi}[$r,D$-ball centered at a point $b$, $B_{D}(b,r)$] $r,D$-ball in $\mathbb R^d$ centered at a point $b$ is a set of points $a$ that are within a $D$-distance of $r$ from $b$:
\begin{equation}
B_{D}(b,r) = \{a \mid a \in \mathbb R^d, D(b,a)\le r\} 
\end{equation}
\label{def:r,d-ball}
\end{defi}

\begin{defi}[Filter] Given a training set of data points $A$, a set of query points $Q$, and a positive real number $r$, the Filter phase outputs all the data points in $A$ that are within a $D$-distance $r$ of each query point $q\in Q$ (i.e. $A\cap B_D(q, r)$).
\label{def:filter}
\end{defi}

\begin{defi}[Refine] Given a natural number $k$ and a set of points in $B_D(q, r)$ for each query point $q\in Q$, the Refine phase outputs the $k$ closest points to $q$ according to the $D$ distance.
\label{def:refine}
\end{defi}

\begin{theorem} [Correctness of Arkade FR reduction]
    Given a training set of data points $A$, a set of query points $Q, q\in Q$, a natural number $k$, a positive real number $r$, and a distance function $D$, Algorithm \ref{alg:arkade} computes the $k$ nearest data points of $q$ within a $D$-distance of $r$ from $q$.
\end{theorem}

\begin{proof}
We first show that any point removed by the {\bf Filter} phase of Algorithm~\ref{alg:arkade} is not inside $B_D(q, r)$.
Then, we show that any point not within the $k$ closest points to $q$ gets removed by the {\bf Refine} phase.
We use these two claims to conclude that the set of points returned by Algorithm~\ref{alg:arkade} is exactly the $k$ nearest neighbors to $q$ within $B_D(q, r)$.

We first claim that the {\bf Filter} phase does not remove any points inside $B_D(q, r)$.
Let $a$ be a point in $A$ and $G_a$ be the AABB centered at $a$.
Notice that by construction, the $r,D$-ball centered at point $a$, $B_D(a, r)$ is contained in the AABB $G_a$ (i.e., $B_D(a, r)\subseteq G_a$).
The point $a$ is removed by the {\bf Filter} phase exactly when the point ray originating from $q$ does not intersect $G_a$, which by the discussion in Section~\ref{sec:$k$NN-on-rt} means $q$ is not a point on or inside $G_a$, so $q$ is not an element of $B_D(a, r)$ which implies that the $D$ distance between $q$ and $a$ is greater than $r$.
\[ q\not\in G_a \implies q\not\in B_D(a, r) \implies D(a, q) > r\]
However, this also implies that $B_D(q, r)$ does not contain $a$ (i.e., $a\not\in B_D(q, r)$) and hence $a$ should be removed. 

Now, we claim that any point not within the $k$ nearest neighbors of $q$ gets correctly removed by the {\bf Refine} phase.
Let $a\in A$ be a point not removed by the {\bf Filter} phase (so $D(a, q) \leq r$) but such that $a$ is not one of the $k$ nearest neighbors to $q$.
This means that there must be $k$ {\em other} points $a_1, a_2, \dots, a_k$ such that the farthest of $k$ neighbors is closer to $q$ than $a$ is (i.e., $a_i\neq a$ and $\max_i D(a_i, q) \leq D(a, q)$).
Then $a$ gets removed on line~\ref{line:12} of Algorithm~\ref{alg:arkade}. 

Since Algorithm~\ref{alg:arkade} does not remove any points that should be kept, and does not keep any points that should be removed, its output is exactly the $k$ points in $B_D(q, r)$ that are closest to $q$. 
\end{proof}

\section{Monotone Transformation}
\label{sec:design-monotone}
This section introduces a new reduction, Arkade Monotone Transformation (MT), that handles some metrics outside $L^p$ better than the Arkade FR reduction.
The $L^p$ distance functions, the primary focus of Section~\ref{sec:filter-refine}, share an important property:
their $r,D$-balls correspond to geometric shapes that can be efficiently represented and processed by RT cores. 
But this property fails for some important distances, e.g. the cosine distance.
To accommodate some of such distances (including cosine), the Arkade MT reduction uses monotone transformations to reduce $k$NN in the given metric to $k$NN in $L^2$.
The resulting $k$NN problem is solved with the well-established $L^2$-distance based RT-accelerated search using spheres~\cite{evangelou}, which is implemented as the $L^2$-instance of Arkade FT.


Arkade MT reduction is based on the following property. 

\begin{defi}[Monotonicity of distance functions]
A distance function $D$ on $\mathbb{R}^n$ is {\em monotonically} {\em increasing} (resp. {\em decreasing}) at a point $q\in\mathbb{R}^n$ if there exists a transformation $f \colon \mathbb{R}^n \to \mathbb{R}^n$ such that for any two points $a_1$ and $a_2$ in $\mathbb R^n$, if $q$ is closer to $a_1$ than $a_2$ in terms of the distance $D$, then after applying the transformation, $f(q)$ is still closer to (resp. further from) $f(a_1)$ than $f(a_2)$ in terms of $L^2$ distance:
\begin{align*}
D(q, a_1) & < D(q, a_2) \Longrightarrow \\
          & L^2(f(q), f(a_1)) < L^2(f(q), f(a_2))\\
\text{(resp. }& L^2(f(q), f(a_1)) > L^2(f(q), f(a_2))
\text{)}.
\end{align*}

A distance function $D$ is {\em monotonically increasing (decreasing)} if it is monotonically increasing (resp. decreasing) at every point $q\in\mathbb{R}^n$.
\label{def:monotone_prop}
\end{defi}

The Arkade MT reduction transforms the input points such that the ordering of the points according to the given distance function is preserved when the transformed points are ordered according to the $L^2$ distance.
The preservation of the ordering can be either positive or negative i.e., the ordering of the transformed points is either the \textit{same} or the \textit{reverse} as that of the original points.
Now, we formally define the Arkade MT reduction in Definition~\ref{def:arkade-mt}.

\begin{defi}[Arkade Monotone Transformation Reduction]
    Given a training set of data points $A$, a set of query points $Q, q\in Q$, a natural number $k$, a monotonic distance metric $D$ and the corresponding transformation $f$, Arkade Monotone Transformation reduction applies $f$ to points in $A$ and $Q$ and performs the Arkade Filter-Refine reduction with $L^2$ distance to find $k$ nearest neighbors of every query point from the set of data points.
    \label{def:arkade-mt}
\end{defi}




\paragraph*{\textbf{Cosine Distance}}

As shown in Figure~\ref{fig:l2-vs-cosine}, the cosine distance between arbitrary vectors does not correlate with the Euclidean distance between vectors' endpoints. 
To introduce a correlation between the cosine and Euclidean distances, the transformation $f$ we apply is normalization. 
In Equation~\ref{equation:fm_cosine}, the normalization multiplier divides each of the coordinate components $a_x$, $a_y$, and $a_z$ of the vector $a$ by the vector's magnitude $\omega$.
\begin{equation}
    f\colon(a_x,a_y,a_z) \to \left(\frac{a_x}{\omega}, \frac{a_y}{\omega}, \frac{a_z}{\omega}\right), \omega=\sqrt{a_x^2+a_y^2+a_z^2}
    \label{equation:fm_cosine}
\end{equation}

When the vectors are normalized, the end points (data points) fall on the unit circle.
The cosine distance between two vectors is the same as the cosine distance between their normalized versions.
Let $\alpha$ be the angle between the query point $q$ and data point $a$. 
Because $q$ and $a$ are normalized, they have a unit magnitude.
The relation between their Euclidean and cosine distances would be the following:
\begin{align*}
L^2(q,a) &= \sqrt{||q||^2 + ||a||^2 - 2 ||q|| ||a|| cos(\alpha) }\\
         &= \sqrt{1^2 + 1^2 - 2\cdot1^2cos(\alpha)} = \sqrt{2 - 2cos(\alpha)}.
\end{align*}

According to the above relation between the Euclidean and cosine distances in the normalized space, 
as the cosine distance between the vectors $q$ and $a$ increases, the angle ($\alpha$) they make at the center decreases, and so the endpoints (data points) of the vectors move closer to the circle, which makes the Euclidean distance between the endpoints smaller.
Therefore, the cosine distance decreases as the Euclidean distance between two data points increases. 
While cosine distance ordering is negatively preserved (reversed) by Euclidean distance ordering, 
the Angular distance ($\alpha$) ordering, is positively preserved by Euclidean distance ordering.

\begin{figure}[!ht]
  \centering
  \includegraphics[width=0.8\linewidth]{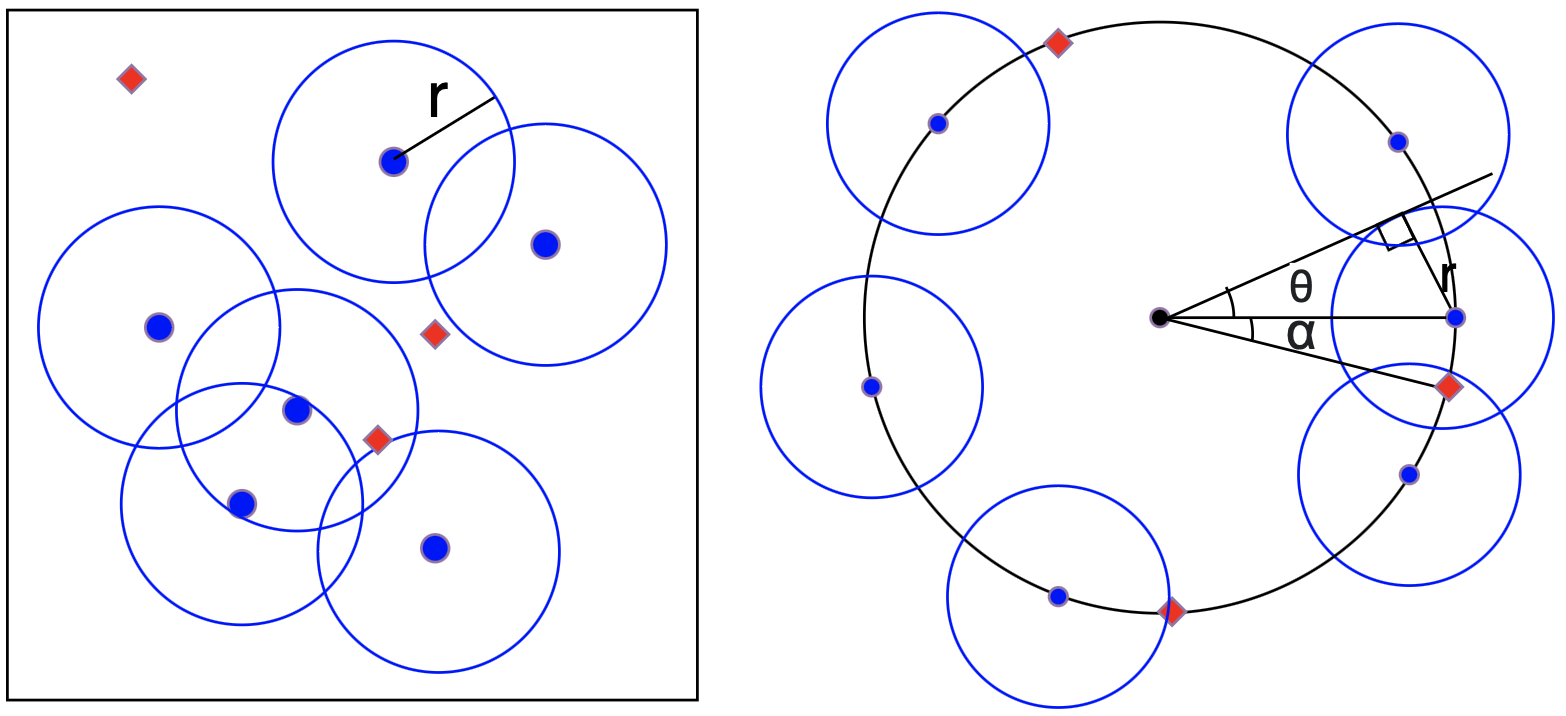}
  \caption{Arkade Monotone Transformation reduction: normalizing points for the cosine distance. Data and query points are marked with, blue and red colors, respectively. $L^2$-based Arkade FR reduction can only be applied after the Monotonic Transformation (normalization) to get the correct cosine distance-based $k$NN.}
  \Description[Arkade Monotone Transformation reduction]{normalizing points for the cosine distance. $L^2$-based Arkade FR reduction can only be applied after the Monotonic Transformation (normalization) to get the correct cosine distance-based $k$NN.}
  \label{fig:cos_redcn}
\end{figure}

In Figure \ref{fig:cos_redcn}, the left and right pictures represent the original and transformed points, respectively.
The left part shows that simply building $L^2$ distance-based spheres and using the Arkade FR reduction with $L^2$ distance will not give us correct neighbors according to the cosine distance. 
The right part signifies that since the normalization preserves the ordering, the Arkade MT reduction can feed the normalized points to the Arkade RF reduction to get the correct $k$ nearest neighbors according to the cosine distance.

\section{Discussion}\label{sec:discussion}
\subsection{Inclusion property to generalize RT-kNN}%
\label{subsec:including-other-distances}

The closest prior work, RT-$k$NN (Section~\ref{sec:$k$NN-on-rt}), is limited to the $L^2$ distance.
In this subsection, we define the inclusion property of a distance function, which allows us to generalize RT-$k$NN to other metrics.
We also explain why this generalization cannot perform better than Arkade and will typically perform worse.

\begin{figure}[h!]
  \centering
  \includegraphics[width=0.6\linewidth]{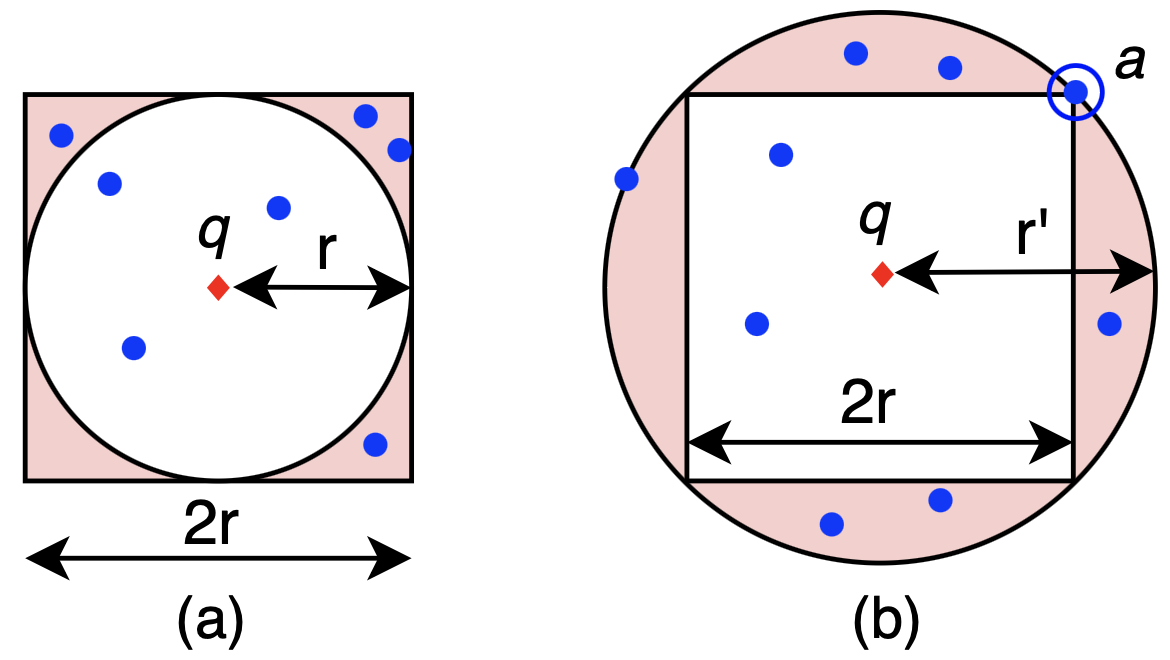}
  \caption{RT-$k$NN reduction can be extended to perform non-$L^2$-based $k$NN with a larger $r$ and $k$ using Inclusion property (Definition~\ref{def:inclusion}).}
  \Description[Extending RT-$k$NN reduction]{RT-$k$NN reduction can be extended to perform non-$L^2$-based $k$NN with a larger $r$ and $k$ using Inclusion property (Definition~\ref{def:inclusion}).}
  \label{fig:extending-l2}
\end{figure}

The RT-$k$NN reduction cannot solve the problem with an arbitrary metric $D$ without certain alterations.
For example, consider the $L^\infty$ distance and the $r$-bounded $k$NN problem.
If we supply the RT-$k$NN reduction with the radius $r$, the candidates outside of the circle but inside the square (an $L^\infty$ ``circle'') will not be found (subfigure~\ref{fig:extending-l2}(a)) and become false negatives.
On the other hand, we could try to supply the RT-$k$NN reduction with a radius $r'$ larger than $r$ (e.g. $r'=\sqrt{2}r$, subfigure~\ref{fig:extending-l2}(b)).
In that case, the $k$ closest neighbors computed according to the $L^2$ distance are not the same as that of the $L^\infty$ distance.
In particular, point $a$ is further than any point in the space between the square and circle according to the $L^2$ distance, but the reverse is true according to the $L^\infty$ distance. Hence, RT-$k$NN reduction may have to exclude point $a$ from the resulting set of nearest neighbors, while the point should be in the set according to the $L^\infty$ metric. Point $a$ becomes a false negative in this case. Note, that this issue can be avoided if we increase $k$ to some $k'$.
In general, by choosing $r'$ and $k'$ arbitrarily larger than the given $r$ and $k$, we can use the RT-$k$NN reduction to perform the $k$NN search based on a non-$L^2$ distance $D$, although it may take extra time to test the candidates that could have been discarded early. 

We call the property of a distance $D$ that allows us to find a finite $r'$ the {\em Inclusion Property} (Definition~\ref{def:inclusion}).
This property states that it is possible to find the $k$NN candidates according to distance $D$ as $k$NN candidates according to $L^2$ distance.
For example, consider the $L^\infty$ distance and $r=1$. 
By the inclusion property, it is possible to construct a finite-sized sphere $S$ of radius $r'=\sqrt{2}$ according to the $L^2$ norm such that all the points that are within a distance $r$ according to $L^\infty$ norm will fall inside sphere $S$.

\begin{defi} [$L^2$-inclusion property of distance $D$]
A distance function $D$ holds the \emph{$L^2$-inclusion property} if for any point $b$ and positive real number $r$  there exists a positive real number $r'$ such that the $r,D$-ball around point $b$ is contained in the $r',L^2$-ball around point $b$.
\begin{equation}
\exists r': B_{D}(b,r) \subset B_{L^2}(b,r').
\end{equation}
\label{def:inclusion} 
\end{defi}

However, the inclusion property is only helpful when the sphere of radius $r'$ efficiently filters the candidates.
In the case of distances where $r'$ is too large that it includes all the points in the dataset, the search process is no better than a linear scan.
Moreover, the inclusion property only addresses how to choose $r'$ but not $k'$ for given $r$ and $k$, respectively.
Currently, we choose $k'$ through trial-and-error. 
We keep on incrementing $k'$ until the RT-$k$NN can find all the $k$ actual nearest neighbors.

\subsubsection{RT-$k$NN reduction vs Arkade FR reduction}
The benefit and utility of Arkade FR reduction over RT-$k$NN reduction can be clearly seen in the case of $L^p$ norms.
With RT-$k$NN, as $p$ increases, $r'$ increases from $r$ to $\sqrt{3}r$ in 3D space\footnote{Exact value of $r'$ would be $max(r, r \cdot d^{\frac{1}{2}-\frac{1}{p}})$. When $p$ is less than $2$, the $r'$ from the inclusion property is the same as the input $r$.}.
The side length of AABBs that contain the corresponding spheres also increases from $2r$ to $2\sqrt{3}r$.
But with Arkade FR reduction, the side length of AABB remains the same at $2r$.
Arkade FR reduction highlights that we do not have to be constricted to using only spheres when we can directly place the distance function geometric objects inside AABB. 
Moreover, the RT cores index the AABBs and not the spheres.

AABBs in Arkade FR reduction are smaller than in the case of RT-$k$NN reduction.
Tighter and smaller AABBs potentially cause less overlap between AABBs, which in turn reduces the number of unnecessary ray-AABB intersections, thus making the reduction run faster.
In $L^p$ norms, as $p$ increases, the geometry of the $L^p$ norm morphs from a sphere into a cube.
Note that a cube is also an AABB.
Therefore, as $p$ increases, the region inside AABB that does not contribute to the $L^p$ norm geometry decreases.
The probability that a ray-AABB intersection would be an output of filter phases increases and reaches $1$ for $L^\infty$ norm.
Hence, for $L^\infty$ norm, a ray-AABB intersection can be forwarded by the Filter phase of Arkade FR reduction without having to perform an additional geometry check since the geometric object and the AABB are the same.
So, Arkade FR reduction achieves optimal performance for a $L^\infty$ norm-based $k$NN search.

\subsection{Other Distances}

\paragraph{Jaccard Distance}
Jaccard similarity($JS$) between two sets $A,B$ is the $A\cap B/ A \cup B$ and Jaccard distance is $1-JS$~\cite{jaccard}. 
The reduction depends on the type of data and the application for which the data is being ranked. 
Assume that set $A$ is represented by a bit vector $Av$ 
where $i^{th}$ bit indicates the presence of $i^{th}$ in the set.
\begin{align*}
JS &= \frac{|A\cap B|}{|A\cup B|}
\end{align*}

Preserving the order according to Jaccard distance and mapping the distance computation is not feasible either using Arkade FR or Arkade MT reductions.
With the existing work on repurposing RT architecture, we do not have a way to perform set operations using RT architecture yet. 
We leave this as a future work.

\paragraph{Hamming Distance}
Given two binary data strings, the Hamming distance is the count of bit positions in which the respective bits of the strings are different~\cite{hamming}.
Hamming distance is equivalent to the Manhattan distance on binary strings.
Indeed, in 3D space, all the possible binary data strings represent vertices of a unit cube, and the Hamming distance between these strings, therefore, is the number of edges that need to be walked from one vertex to the other.
Hence, we can use $L^1$ norm-based Arkade FR reduction.

\paragraph{Mahalanobis Distance}
Mahalanobis distance is the distance between a point and a given distribution, where the standard deviation of the point is compared to the mean of the distribution.
After a particular spatial transformation, when the axes are scaled to unit variance, Mahalanobis distance is Euclidean distance \cite{mahalanobis}.
Hence, we can use $L^2$ norm-based Arkade FR reduction.

\subsection{Choice of radius}
Our reduction require radius ($r$) as an input parameter. 
Selecting a {\em optimal} radius is a challenging task because an arbitrary choice of radius might result in poor performance or accuracy.
However, this issue is orthogonal to Arkade's reductions. 
Presently, we adopt an approach of prior work, TrueKNN~\cite{trueknn}, which we elaborate on in the Evaluation Section~\ref{sec:eval-setup}. 
Exploring alternative approaches for determining optimal radius is an intriguing avenue for future investigation.



\section{Evaluation}\label{sec:eval}
In this section, we evaluate Arkade's Filter\hyp{}Refine and Monotone Transformation reductions on four groups of realistic datasets using four baselines. 
We analyze various factors such as the BVH tree quality, the average number of ray-AABB intersections, and the number of rounds (defined in Sec.~\ref{sec:eval-num-rounds}) that impact the performance of Arkade on these datasets.
Then, we look at the effect of parameters such as $k$.

\subsubsection*{\textbf{Datasets}}

The characteristics of the datasets we used are summarized in Table~\ref{table:datasets}. As RT cores can only build BVH on three-dimensional data, we use only 2D and 3D datasets. For 2D datasets, we set the third dimension to zero.

\begin{description}
    \item[Geospatial Datasets] 
    Gowalla~\cite{gowalla} dataset contains check-in locations of users from across the world in the form of latitude and longitude~\cite{gowalla}. We processed the dataset to get only distinct locations.
    Cali OSM~\cite{osm} contains geo-spatial coordinates of a very small region in California, sourced from OpenStreetMap. Because the coordinates are local, we treat it as 2-dimensional data.
    Gbif~\cite{gbif} contains information on several birds and the locations where they are spotted. We obtained the geospatial coordinates of the spottings for January 2018.
    We convert the geospatial coordinates into Cartesian coordinates before passing the data as input to the Arkade reductions.
    
    \item[Point Clouds] Kitti~\cite{kitti} is an autonomous driving footage popularly used in computer vision benchmarks. The data we used is in the form of 3D point clouds generated by the Velodyne scanner. We combined several frames to make up our dataset.
    Randnet~\cite{Chen_2022_BMVC} is a synthetic point cloud generated from real-world and synthetic environments using RandLA-Net architecture~\cite{hu2019randla}.
    This particular dataset is built on an aerial view of a city landscape.
    
    \item[3D Scans] Manuscript~\cite{manuscript} dataset is an XYZ RGB 3D scan of a page in Latin from Vellum manuscript. 

    \item[Synthetic Datasets] Glove 3D is a three-dimensional PCA projection of 25-dimensional Glove data~\cite{glove}.
    Randnet is also a synthetic dataset.
\end{description}

\begin{table}[h]
\centering
\caption{Datasets Characteristics}
\label{table:datasets}
\begin{tabular}{lcccl}\toprule
 Dataset & Data Points & Queries & Dimension \\\midrule
 Gowalla~\cite{gowalla} & 1270969 & 10000 & 3 \\ 
 Glove 3D~\cite{glove} & 1183514 & 10000 & 3 \\
 Manuscript~\cite{manuscript} & 2145617 & 10000 & 3 \\ 
 Cali OSM~\cite{osm} & 4195951 & 10000 & 2 \\ 
 Kitti~\cite{kitti} & 4000000 & 10000 & 3 \\ 
 Randnet~\cite{hu2019randla} & 6815065 & 10000 & 3 \\ 
 Gbif~\cite{gbif} & 8475714 & 10000 & 3 \\\bottomrule 
\end{tabular}
\end{table}

\subsubsection*{\textbf{Baselines}}
We used three GPU and one state-of-the-art CPU $k$NN libraries to evaluate Arkade. This mixture contains both tree-based and non-tree-based approaches.

\begin{description}
    
    \item[SCANN] is a quantization-based {\em approximate} similarity search library~\cite{scann}. 
    It is the state-of-the-art in CPU $k$NN implementations~\cite{ann}.
    We use the same parameters as ANN benchmarks~\cite{ann} to get a recall\footnote{Recall is the ratio of the number of correctly found nearest neighbors by the search to the number of true nearest neighbors from the ground truth.} of 0.99.
    
    \item[Treelogy] implements a KD-tree-based {\em exact} GPU implementation~\cite{autoropes}.
    We modify the Treelogy code to perform $L^p$ and cosine distance-based $k$NN search.
    
    \item[FAISS] is a state-of-the-art {\em exact} quantization-based GPU library~\cite{faiss,ann}. 
    FAISS uses tensorflow-gpu to interface with CUDA cores. 
    We use the IVFFlatL2 index (as used in ANN benchmarks~\cite{ann}) and train the data before the search.
    
    \item[FastRNN] uses RT architecture to perform fixed-radius search, only in case of Euclidean distances~\cite{evangelou}. To correctly perform the $k$NN search using other distances, we use a larger radius $\sqrt{d}r$, where $r$ is the given radius and $d$ is the data dimension, and a larger number of nearest neighbors $k'$ just enough to obtain $k$ nearest neighbors according to a given distance. (see Subsection~\ref{subsec:including-other-distances}).
    
\end{description}   

We use Treelogy and FastRNN to evaluate the Arkade Filter-Refine reduction, while we use SCANN, FAISS, and Treelogy to evaluate the Arkade Monotone Transformation reduction.
SCANN and FAISS implement only $L^2$ and cosine distances on CPU and GPU respectively.
On the other hand, the modifications of FastRNN only work for $L^p$ distances.

\subsubsection*{\textbf{Experimental Setup}} \label{sec:eval-setup}
We used NVIDIA GeForce RTX 4070 Ti GPU with 12GB memory for all of our experiments. 
To interface with the RT architecture on the GPU, we used Optix Wrapper Library~\cite{owl}.
Arkade builds the BVH tree index once over the entire set of data points for chosen parameters and searches for neighbors once for all the query points in every run.
We perform 5 such runs to collect and average the performance metrics such as build time and search time.  
All the reported numbers are rounded to two non-zero decimals.

We evaluate Filter\hyp{}Refine reduction with the $L^1$ and $L^\infty$ distance functions, and Monotone Transformation reduction with cosine distance.
We plug in TrueKNN's~\cite {trueknn} approach of choosing a small radius and iteratively increasing the radius until all the query points find their $k$ neighbors.
To make a fair comparison, we also apply TrueKNN to the baseline FastRNN.

\begin{table*} 
    \centering
    \caption%
      {Search times and speedups of Arkade over all baselines for distance functions $L^1$ and $L^\infty$ }%
    \label{table:speedups}
    \begin{tabular}{l@{}rrr rr  rrr rr} 
        \toprule
        \multirow{3}[3]{*}{Dataset}
          & \multicolumn{5}{c}{$L^1$ distance} & \multicolumn{5}{c}{$L^\infty$ distance} \\
        \cmidrule(lr){2-6}\cmidrule(lr){7-11} 
        
        & \multicolumn{3}{c}{\centering Search time (seconds)} & \multicolumn{2}{c}{Arkade speedup over} & \multicolumn{3}{c}{Search time (seconds)} & \multicolumn{2}{c}{Arkade speedup over}  \\
        \cmidrule(lr){2-4}\cmidrule(lr){5-6}\cmidrule(lr){7-9}\cmidrule(lr){10-11} 
        
        & \multicolumn{1}{c}{Treelogy} & \multicolumn{1}{c}{FastRNN} & \multicolumn{1}{c}{Arkade} & \multicolumn{1}{c}{Treelogy}  & \multicolumn{1}{c}{FastRNN} & \multicolumn{1}{c}{Treelogy} & \multicolumn{1}{c}{FastRNN} & \multicolumn{1}{c}{Arkade} & \multicolumn{1}{c}{Treelogy}  & \multicolumn{1}{c}{FastRNN}  \\\midrule

        Gowalla & 0.16  & 1.06 & 0.032 & \textbf{5.0} & \textbf{33.1} & 0.16 & 0.34  &  0.022 & \textbf{7.3} & \textbf{15.4} \\
        Glove3D & 0.16  & 0.14 & 0.0063 & \textbf{25.4} & \textbf{22.2} & 0.14 & 0.011 & 0.0025 & \textbf{56.0} & \textbf{4.4} \\
        Manuscript & 0.18 & 0.075 & 0.011 & \textbf{16.4} & \textbf{6.8} & 0.19 & 0.032 & 0.010 & \textbf{19.0} & \textbf{3.2}  \\
        CaliOSM & 0.25 & 0.21 & 0.16 & \textbf{1.6} & \textbf{1.3} & 0.24 & 0.11 & 0.029 &\textbf{ 8.3} & \textbf{3.8} \\
        Kitti4M & 0.25 & 0.41 & 0.045 & \textbf{5.6} &\textbf{9.1} & 0.26 & 0.18 & 0.043 & \textbf{6.1}  & \textbf{4.2} \\
        Randnet &0.37  & 0.07 & 0.0023 & \textbf{160.9} & \textbf{30.4}& 0.36 & 0.028 & 0.0018 & \textbf{200.0} &  \textbf{15.6} \\
        Gbif & 0.39  & 1.25 & 0.082 & \textbf{4.8} & \textbf{15.2} & 0.37 & 1.14 & 0.077 & \textbf{4.8} &  \textbf{14.8} \\\bottomrule
    \end{tabular}
\end{table*}

\begin{table*}
    \centering
    \caption%
    {Search times and speedups of Arkade over all baselines for cosine distance function}%
    \label{table:dot_speedups}
    \begin{tabular}{l@{}rrrr@{\hskip 7mm}rrr} 
        \toprule
        
        \multirow{2}[3]{*}{Dataset}
          & \multicolumn{4}{c}{Search time (seconds)} & \multicolumn{3}{c}{Arkade speedup over}  \\
        \cmidrule(lr){2-5}\cmidrule(lr){6-8} 
        
          & \multicolumn{1}{c}{SCANN} & 
          \multicolumn{1}{c}{Treelogy} & 
          \multicolumn{1}{c}{FAISS} & 
          \multicolumn{1}{c}{Arkade} &  
          \multicolumn{1}{c}{SCANN}  & 
          \multicolumn{1}{c}{Treelogy} &
          \multicolumn{1}{c}{FAISS} \\\midrule

        Gowalla & 79.37 & 0.29 & 0.22 & 0.10 & \textbf{793.7} & \textbf{2.9} & \textbf{2.2} \\
        Glove3D  & 76.52 & 0.32 & 0.21 & 0.0033 & \textbf{23,187.9} & \textbf{97.0} & \textbf{63.6}\\
        Manuscript  &  149.04 & 0.47 & 0.38 & 0.012 & \textbf{12,420.0} & \textbf{39.1} & \textbf{16.7}\\
        CaliOSM  & 284.67 & 1.02 & 0.74 & 0.013 & \textbf{21,897.7} & \textbf{78.5} & \textbf{56.9}\\
        Kitti4M & 261.39 & 0.86 & 0.7 & 0.14 & \textbf{1,867.1} & \textbf{6.1} & \textbf{5.0} \\
        Randnet & 471.12 & 1.45 & 1.2 & 0.026 & \textbf{18,120.0} & \textbf{55.8} & \textbf{46.2}\\
        Gbif & 581.01 & 1.81 & 1.49 & 0.29 & \textbf{2,003.5} & \textbf{6.2} & \textbf{5.1} \\
        \bottomrule
    \end{tabular}
\end{table*}

\subsection{Performance Evaluation}
We compare the search times and the speedups of Arkade reductions over all the baselines and the datasets in Tables~\ref{table:speedups} and~\ref{table:dot_speedups}.
Table~\ref{table:speedups} shows the comparison of Arkade to the baselines, Treelogy and FastRNN, for $L^1$ and $L^\infty$ norms.
In Table~\ref{table:dot_speedups}, we show the same performance numbers for Cosine distance.

Among all the baselines, we see that Arkade is significantly faster than SCANN, although the speedup can be attributed to SCANN being a purely CPU-based implementation.
In the case of GPU baselines, Arkade is still faster by $1.5$x-$200$x.
The speedup of Arkade over non-RT baselines demonstrates the ability of RT cores to efficiently accelerate the irregular tree traversals.
The speedups over the RT baseline, FastRNN, show how Arkade efficiently utilizes the RT cores to accelerate a broader range of applications.

In general, we find that the speedups of Arkade over baselines do not increase with an increase in the dataset size.
For example, Gowalla and Glove3D datasets are roughly 1M in size but Arkade's speedups on these datasets are very different.
The search times of non-RT-based implementations such as SCANN, Treelogy, and FAISS increase with the increase in the size of the dataset, however, RT implementations such as Arkade and FastRNN do not follow the same trend.
We go into more detail in Section~\ref{sec:eval-breakdown}.

\subsubsection{$L^1$ norm}
In the first half of Table~\ref{table:speedups}, we see that Arkade achieves speedups of $1.6$x-$160.9$x and $1.3$x-$33.1$x over Treelogy and FastRNN, respectively.
Arkade is faster than Treelogy since Arkade uses RT cores to accelerate the BVH tree traversals, while Treelogy uses shader cores.

In this experiment, we use the same search radius for FastRNN and Arkade.
This is because the $L^1$ norm geometric object (rhombus) is present inside the $L^2$ norm geometric object (circle). 
Even though the search radius is the same, FastRNN searches for a larger number of neighbors. 
FastRNN uses $L^2$ distance to rank the neighbors unlike Arkade, which uses the \textit{actual} distance function, $L^1$ norm.
Because $L^1$ norm geometric object is smaller in volume compared to $L^2$ norm geometric object, Arkade can efficiently search neighbors in a smaller space, which is why Arkade is consistently faster than FastRNN.

\subsubsection{$L^\infty$ norm}
In the second half of Table~\ref{table:speedups}, we see that Arkade achieves speedups of $4.8$x-$200$x and $3.2$x-$15.6$x over Treelogy and FastRNN, respectively.
We find that Arkade outperforms Treelogy for the same reason as in the case of the $L^1$ norm.

As noted in Section~\ref{subsec:including-other-distances}, FastRNN needs a larger search radius ($\sqrt{3}$ times Arkade's radius) and $k$ compared to Arkade. 
When the radius increases, the size of the AABB increases, which causes an increase in the number of ray-AABB intersection tests performed during the BVH traversal.
As these intersection tests are the most computationally intensive part of the ray tracing pipeline, we find that Arkade is significantly faster than FastRNN.
We further analyze the performance of Arkade and FastRNN in Section~\ref{sec:eval-analysis}.

\subsubsection{Cosine distance}
In Table~\ref{table:dot_speedups}, we see that Arkade achieves speedups of 
$793.7$x-$23,187.9$x, $2.9$x-$97.0$x, and $2.2$x-$63.6$x over the baselines, SCANN, FAISS, and Treelogy, respectively.
Though FAISS is the current state-of-the-art GPU-based $k$NN search, it is designed for higher dimensional $k$NN and uses a heavy tensorflow framework.
We believe that the combination of FAISS's overheads and Arkade's RT-accelerated neighbor search algorithm results in Arkade's better performance. 

\subsection{Performance Analysis} \label{sec:eval-analysis}

The speedup trend of Arkade can be explained by the data distribution of the dataset.
This is because the way the data is distributed affects the quality of the constructed BVH, the number of ray-AABB tests performed for each query point, and, consequently, the number of candidates the Filter phase forwards to the Refine phase.
We unroll the effects of data distribution on each of the reductions in the following subsections.

\subsubsection{Breakdown} \label{sec:eval-breakdown}
\begin{figure}[h!]
    \centering
    \begin{minipage}{\linewidth}
        \centering
        \begin{subfigure}{0.49\linewidth}
            \includegraphics[width=\linewidth]{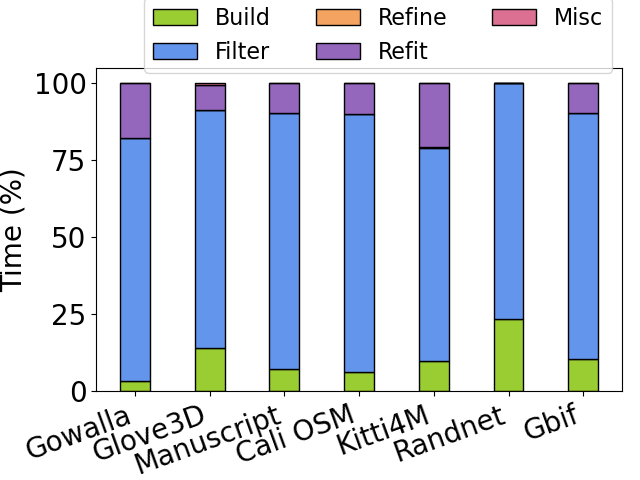}
            \caption{Arkade}
            \label{fig:arkade_break}
        \end{subfigure}
        \hfill
        \begin{subfigure}{0.49\linewidth}
            \includegraphics[width=\linewidth]{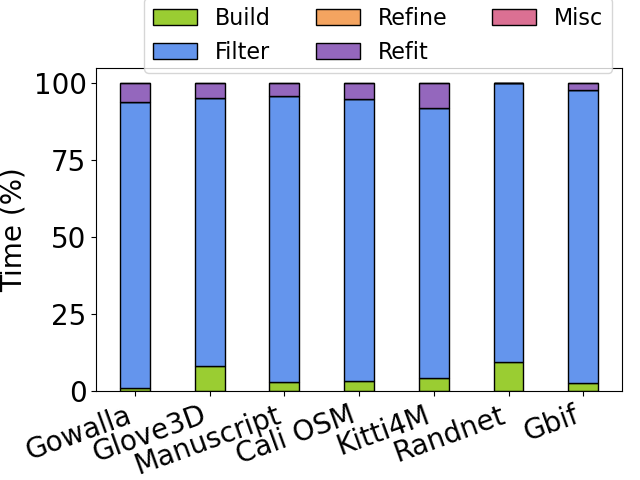}
            \caption{FastRNN}
            \label{fig:fastrnn_break}
        \end{subfigure} 
    \end{minipage}
    \caption{Run time breakdown of RT implementations for $L^\infty$ norm-based $k$NN search}
    \Description[Run time breakdown of RT implementations for $L^\infty$ norm-based $k$NN search]{refer to sections ~\ref{sec:eval-breakdown},\ref{sec:eval-impact-bvh}, \ref{sec:eval-impact-aabb}, and \ref{sec:eval-num-rounds} to understand what this figure conveys.}
    \label{fig:breakdown}
\end{figure}

To understand the factors impacting the Arkade speedups, we present a complete breakdown of Arkade and FastRNN execution times for $L^\infty$ norm in Figure~\ref{fig:breakdown}.
The execution time is comprised of both BVH build and search times.
The search time is further divided into four parts -- time taken by the Filter phase, Refine phase, refit, and miscellaneous maintenance in between these steps.

Figures~\ref{fig:arkade_break} and ~\ref{fig:fastrnn_break} show the breakdown of execution times of Arkade and FastRNN for $L^\infty$ distance, respectively.
The percentage of build time is higher in the case of Arkade than in FastRNN. 
However, the actual build times in both cases are approximately the same for respective datasets.
Because Arkade's search times are lower than FastRNN's, the percentage of build time of Arkade is higher.

In Figures~\ref{fig:arkade_break} and \ref{fig:fastrnn_break}, the Filter phase predominantly takes more time than any other steps.
In the Filter phase, the ray traverses the BVH and checks if it intersects an AABB, and when it does intersect an AABB, it further checks if the ray intersects the geometry.
The time the filter phase takes is affected by the quality of BVH the RT architecture constructs.
The structure of BVH further impacts the BVH traversal and the number of intersection tests performed.

\subsubsection{Impact of BVH Tree Quality}
\label{sec:eval-impact-bvh}
The negligible amount of time spent in the Refine phase (Refine time is barely visible in Figure~\ref{fig:breakdown})  supports the observation that more time is spent in traversal and filtering AABBs rather than ordering the candidates present inside them.
The mapping of the $k$NN problem to RT architecture needs the geometric objects to overlap to produce results.
As the $k$ value increases, the search radius needed to find all $k$ neighbors also increases. 
This increases the potential of geometric objects to overlap and reduces the effectiveness of the BVH in pruning large parts of the neighbor search space, resulting in a BVH of poor quality.
High overlap, in turn, increases the number of ray-AABB intersections.

\begin{table*} 
    \centering
    \caption%
    {Average number of ray-AABB intersection and number of rounds for $L^\infty$ distance (Table~\ref{table:speedups}) and cosine distance (Table~\ref{table:dot_speedups}).}%
    \label{table:metadat}
    \begin{tabular}{l@{}rr rr rr} 
        \toprule
        
        \multirow{4}[3]{*}{Dataset}
          & \multicolumn{4}{c}{$L^\infty$ distance}  & \multicolumn{2}{c}{Cosine distance} \\
        \cmidrule(lr){2-5}\cmidrule(lr){6-7} 
        
        & \multicolumn{2}{c}{Arkade} & \multicolumn{2}{c}{FastRNN} & \multicolumn{2}{c}{Arkade}   \\
        \cmidrule(lr){2-3}\cmidrule(lr){4-5}\cmidrule(lr){6-7} 
        
        & \multicolumn{1}{c}{Average}  &
          \multirow{2}{*}{{Rounds}} &
          \multicolumn{1}{c}{Average}  &
          \multirow{2}{*}{{Rounds}} &
          \multicolumn{1}{c}{Average}  &
          \multirow{2}{*}{{Rounds}} \\

        & \multicolumn{1}{c}{\#Intersections} &
          &
          \multicolumn{1}{c}{\#Intersections} &
          &
          \multicolumn{1}{c}{\#Intersections} &
          \\\midrule
    
        Gowalla & 263.47 & 10  & 
        510.40 & 10 &   
        10613.80 & 7  \\
        Glove3D & 26.12 & 2 & 
        60.46 & 2 &    
        60.46 & 2  \\
        Manuscript & 173.20 & 4 & 
        510.50 & 4 &    
        357.77 & 3 \\
        CaliOSM & 440.11 & 6 & 
        827.03  & 6 &     
        2695.24 & 1  \\
        Kitti4M & 366.38 & 8  & 
        365.28  & 7 &    
        20669.20 & 1 \\
        Randnet & 121.92 & 1  &  
        397.79 & 1 &  
        211.86 & 4 \\
        Gbif & 3093.40 & 5 &  
        5185.58 & 5 &  
        20708.30 & 1  \\
        \bottomrule
    \end{tabular}
\end{table*}

\subsubsection{Impact of ray-AABB intersections} \label{sec:eval-impact-aabb}

In Table~\ref{table:metadat}, we present the average number of ray-AABB intersections per query point that occurred in the RT-based implementations in the case of $L^\infty$ and Cosine distances.
In the case of Cosine distance, the search times of Arkade increase with an increase in the number of intersections and decrease with a decrease in the number of intersections.
Similarly, in the case of $L^\infty$ distance, Arkade and FastRNN search times are proportional to the number of intersections except for the Kitti4M dataset.
Moreover, the number of intersections is higher for any dataset in the case of Cosine distance compared to $L^\infty$ norm, and we observe that Cosine distance-based search takes longer than that of $L^\infty$ norm.
Due to normalization, the points become denser in the Cosine distance scenario. 

\subsubsection{Impact of number of rounds} \label{sec:eval-num-rounds}

While the number of intersections explains most of the trends in search times of RT-based implementations, there are certain instances where the number of intersections alone does not suffice.
For example, FastRNN spends more search time on Kitti4M than the Randnet dataset, but the number of ray-AABB intersections on Kitti4M is lower than that of Randnet.
We observe that the number of rounds is higher in the case of Kitti4M than in Randnet.
We present the number of rounds for each RT-implementation and dataset in Table~\ref{table:metadat}.
The number of rounds is the number of times TrueKNN doubles the radius until it finds $k$ nearest neighbors of all query points.

A higher number of rounds increases the refit time.
In Figure~\ref{fig:breakdown}, we see that refit is the second most time-consuming part of the search. 
The refit time corresponds to doubling the radius of geometries, updating the AABBs to fit the new larger geometries, and refitting the BVH for every round. 

The need for a higher number of rounds arises from the data distribution. 
A new round is performed when the neighbors of some of the query points can only be found at a larger radius.
Hence, the number of rounds indicates that some neighborhoods of the dataset are denser than others.

\subsubsection{Sensitivity to $k$}

\begin{figure}[h!]
    \centering
    \begin{minipage}{\linewidth}
        \centering
        \begin{subfigure}{0.49\linewidth}
            \includegraphics[width=\linewidth]{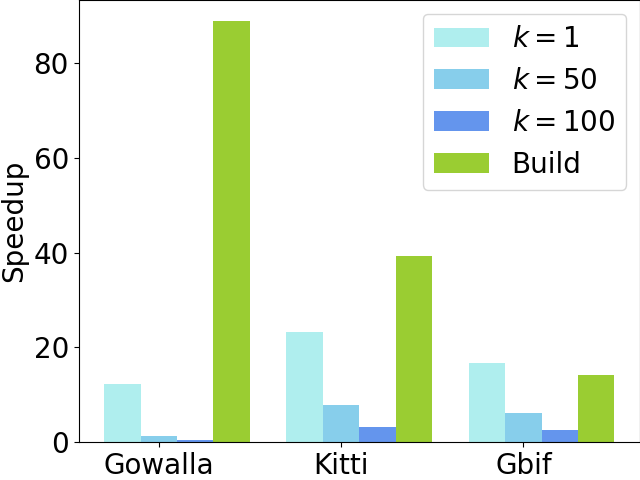}
            \caption{For $k=1,50,100$, Arkade's search and build time speedup over Faiss}
            \Description[Arkade's search and build time speedup over Faiss with Cosine distance]{For $k=1,50,100$}
            \label{fig:dot_k}
        \end{subfigure}
        \hfill
        \begin{subfigure}{0.49\linewidth}
            \includegraphics[width=\linewidth]{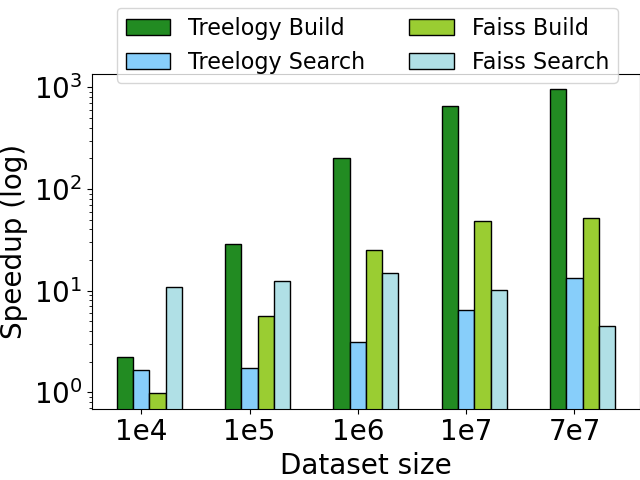}
            \caption{Arkade's search and build time speedups over Faiss and Treelogy as the dataset size changes}
            \Description[Arkade's search and build time speedup over Faiss with Cosine distance]{For $k=1,50,100$}
            \label{fig:dot_mag}
        \end{subfigure} 
    \end{minipage}
    \caption{Sensitive analysis of Arkade's search and build time speedups for Cosine distance}
    \Description[Sensitive analysis of Arkade's search and build time speedups for Cosine distance]{speedups as k and magnitude of dataset size change}
    \label{fig:justarkade}
\end{figure}

In Figure~\ref{fig:dot_k}, we study the speedup of Arkade performance over FAISS in the case of Cosine distance as $k$ increases.
We vary from $k$ as 1, 50, and 100 on Gowalla, Kitti4M, and Gbif datasets.
We also plot the build time speedup of Arkade over FAISS for each dataset.
Arkade's search time speedups decrease as $k$ increases.
But, observe that the build times of Arkade are much lower than FAISS. 
So the overall runtime (build + search time) of Arkade is still lower than FAISS.
At a sufficiently large dataset size and $k$, it is possible that the benefit of using Arkade might diminish.
However, in practice, $k$ is typically at most $100$~\cite{ann}.

\subsubsection{Sensitivity to Dataset size}

In Figure~\ref{fig:dot_mag}, we study the speedup (in log scale) of Arkade performance over Treelogy and FAISS in the case of Cosine distance as the magnitude of the dataset size increases.
We uniformly sample a randomly generated dataset to get the number of data points from 10K to 70M.
In Figure~\ref{fig:dot_mag}, green shaded bars show the speedup of Arkade build times over Treelogy and Faiss, respectively, while blue shaded bars show the speedup of Arkade search times over the same baselines.
Arkade's build and search time speedups over Treelogy increase almost linearly with the increasing magnitude of the dataset size. 
We attribute these speedups to the optimized strategies employed for tree construction and traversal by the RT architecture.
Conversely, search time speedups over Faiss slightly decrease when the dataset size reaches 10M.
We also note that Arkade runs out of memory on our 12GB GPU after 70M points, however, both the baselines can execute up to 100M points.

\subsubsection{Impact of hardware utilization}
Available Nvidia profilers~\cite{nsight} can not differentiate RT cores from shader cores. 
Nvidia 4060Ti, the GPU on which we ran our experiments, has 4352 and 32 shader and RT cores, respectively. 
Without knowing how the architecture maps the point ray queries to the hardware, it is difficult to determine if Arkade is saturating the resources on RT architecture.
However, applying optimizations like balancing the workload among the threads may improve the utilization and performance.
We leave this for future work.

\section{Related Work}
\subsection{Non-RT Applications Accelerated With RT Architecture}
\label{sec:relatedrt}

Recent work has shown that non-ray-tracing problems can be expressed as ray-object intersection problems, making them amenable to acceleration with RT cores~\cite{wald, zellman, evangelou,rtnn, dbscan}. 
Wald \textit{et. al.}~\cite{wald} were the first to use RT cores to accelerate non-ray tracing applications. They looked at the problem of identifying the location of a point in a tetrahedral mesh. By modeling the point as a ray and reporting the closest tetrahedron intersected by the ray, they identified the tetrahedron in which the point was contained. 
Zellman \textit{et. al.}~\cite{zellman} showed how to use RT cores to perform graph drawing. They re-formulated the nearest neighbor search subroutine as a ray tracing problem and used the force exerted by the nearest neighbors to direct their graph drawing algorithm. They found their approach to be significantly faster than the state-of-the-art force-directed graph drawing algorithms.
Evangelou \textit{et. al.}~\cite{evangelou} used RT cores to perform photon mapping by finding the set points in a fixed-radius neighborhood of a query point. They used the reduction proposed by Zellman \textit{et. al.} and found that they were up to 15x faster than non-RT-accelerated baselines.
Zhu \textit{et. al.}~\cite{rtnn} proposed optimizations such as point reordering and query partitioning to improve the performance of RT-accelerated neighbor searches.  
Nagarajan \textit{et. al.} proposed RT-DBSCAN~\cite{dbscan} and TrueKNN~\cite{trueknn} to leverage RT cores to solve DBSCAN clustering and efficiently perform $k$-nearest neighbor search, respectively.

\subsection{Tree-based, GPU-accelerated kNN}

Tree-based $k$NN algorithms are only efficient at lower dimensions due to the curse of dimensionality~\cite{vafile}. They are mostly specialized for certain applications.
Merry \textit{et. al.}~\cite{kdtree-knn} propose an optimization to leverage the coherence of points when traversed in $k$d tree order so as to reuse traversal information of neighboring points. They find that their approach is 4.4x to 4.6x faster and does not require any modifications to the $k$d tree. 
Treelogy\cite{treelogy, autoropes} proposes several optimizations to improve memory coalescing and reduce divergence caused by GPU threads that traverse different parts of the tree.
Gieseke \textit{et. al.}~\cite{bufferkdtree} propose the idea of a buffer $k$d tree to create batches of query points that all target the same leaf nodes of the $k$d tree, exploiting data locality. However, their work is specialized for data with dimensionality between 4 and 25.
An optimized approximate KD-tree-based KNN is proposed to aid in point cloud registration~\cite{tigris}. However, this optimization is application-specific.
Gowanlock \textit{et. al.}~\cite{gowanlock2021hybrid} proposes a hybrid CPU-GPU algorithm that breaks computation up so that areas of large density are assigned to the GPU, while the CPU handles the rest of the data. This approach leverages the advantages offered by the different architectures to optimize performance.

\section{Conclusion}
Irregular problems like tree traversals are ubiquitous, especially queries like nearest neighbor search that have applications in domains such as point cloud registration in computer vision, data compression, similarity scoring, DNA sequencing, etc.
Tree-based nearest neighbor search is naturally challenging to scale up using purely software approaches on massively parallel commodity hardware such as GPUs.  
Even though ray tracing cores of GPU are specialized hardware to cater to graphics applications,
we show that this specialized hardware can be generalized to accelerate tree operations in other domains,
To that end, we provide a set of reductions to the ray tracing scene.
Without our reductions, distance metric computations such as $L^p$ norm and cosine distance take significantly longer to complete or cannot be run on RT cores (it varies between previous works). 
While RT cores accelerate tree traversals through BVH construction, this tree structure is not accessible to the user and is limited to 3D space. 
Availability and programmability of the spatial tree itself would be more helpful in using RT cores for general applications.

\begin{acks}
We are thankful to all the anonymous reviewers for providing valuable feedback. We also thank Kirshanthan Sundararajah for helping us improve the earlier versions of the paper and Raghav Malik for helping us with the proof. This work was funded by NSF grants CCF-1908504, CCF-1919197 and CCF-2216978.
\end{acks}

\bibliographystyle{ACM-Reference-Format}
\bibliography{references}


\begin{thebibliography}{54}


\ifx \showCODEN    \undefined \def \showCODEN     #1{\unskip}     \fi
\ifx \showDOI      \undefined \def \showDOI       #1{#1}\fi
\ifx \showISBNx    \undefined \def \showISBNx     #1{\unskip}     \fi
\ifx \showISBNxiii \undefined \def \showISBNxiii  #1{\unskip}     \fi
\ifx \showISSN     \undefined \def \showISSN      #1{\unskip}     \fi
\ifx \showLCCN     \undefined \def \showLCCN      #1{\unskip}     \fi
\ifx \shownote     \undefined \def \shownote      #1{#1}          \fi
\ifx \showarticletitle \undefined \def \showarticletitle #1{#1}   \fi
\ifx \showURL      \undefined \def \showURL       {\relax}        \fi
\providecommand\bibfield[2]{#2}
\providecommand\bibinfo[2]{#2}
\providecommand\natexlab[1]{#1}
\providecommand\showeprint[2][]{arXiv:#2}

\bibitem[Adeniyi et~al\mbox{.}(2016)]%
        {adeniyi2016automated}
\bibfield{author}{\bibinfo{person}{David~Adedayo Adeniyi},
  \bibinfo{person}{Zhaoqiang Wei}, {and} \bibinfo{person}{Yang Yongquan}.}
  \bibinfo{year}{2016}\natexlab{}.
\newblock \showarticletitle{Automated web usage data mining and recommendation
  system using K-Nearest Neighbor (KNN) classification method}.
\newblock \bibinfo{journal}{\emph{Applied Computing and Informatics}}
  \bibinfo{volume}{12}, \bibinfo{number}{1} (\bibinfo{year}{2016}),
  \bibinfo{pages}{90--108}.
\newblock


\bibitem[AMD(2023)]%
        {amd}
\bibfield{author}{\bibinfo{person}{AMD}.} \bibinfo{year}{2023}\natexlab{}.
\newblock \bibinfo{title}{AMD Ray tracing}.
\newblock
\newblock
\urldef\tempurl%
\url{https://www.amd.com/en/technologies/rdna}
\showURL{%
\tempurl}


\bibitem[Andr{\'{e}} et~al\mbox{.}(2015)]%
        {andre2016cache}
\bibfield{author}{\bibinfo{person}{Fabien Andr{\'{e}}},
  \bibinfo{person}{Anne{-}Marie Kermarrec}, {and} \bibinfo{person}{Nicolas~Le
  Scouarnec}.} \bibinfo{year}{2015}\natexlab{}.
\newblock \showarticletitle{Cache locality is not enough: High-Performance
  Nearest Neighbor Search with Product Quantization Fast Scan}.
\newblock \bibinfo{journal}{\emph{Proc. {VLDB} Endow.}} \bibinfo{volume}{9},
  \bibinfo{number}{4} (\bibinfo{year}{2015}), \bibinfo{pages}{288--299}.
\newblock


\bibitem[Aumüller et~al\mbox{.}(2020)]%
        {ann}
\bibfield{author}{\bibinfo{person}{Martin Aumüller}, \bibinfo{person}{Erik
  Bernhardsson}, {and} \bibinfo{person}{Alexander Faithfull}.}
  \bibinfo{year}{2020}\natexlab{}.
\newblock \showarticletitle{ANN-Benchmarks: A benchmarking tool for approximate
  nearest neighbor algorithms}.
\newblock \bibinfo{journal}{\emph{Information Systems}}  \bibinfo{volume}{87}
  (\bibinfo{year}{2020}), \bibinfo{pages}{101374}.
\newblock
\showISSN{0306-4379}
\urldef\tempurl%
\url{https://doi.org/10.1016/j.is.2019.02.006}
\showDOI{\tempurl}


\bibitem[Bentley(1975)]%
        {bentley1975multidimensional}
\bibfield{author}{\bibinfo{person}{Jon~Louis Bentley}.}
  \bibinfo{year}{1975}\natexlab{}.
\newblock \showarticletitle{Multidimensional binary search trees used for
  associative searching}.
\newblock \bibinfo{journal}{\emph{Commun. ACM}} \bibinfo{volume}{18},
  \bibinfo{number}{9} (\bibinfo{year}{1975}), \bibinfo{pages}{509--517}.
\newblock


\bibitem[Bourbaki(1987)]%
        {bourbaki1987topological}
\bibfield{author}{\bibinfo{person}{N. Bourbaki}.}
  \bibinfo{year}{1987}\natexlab{}.
\newblock \bibinfo{booktitle}{\emph{Topological Vector Spaces: Chapters 1-5}}.
\newblock \bibinfo{publisher}{Springer-Verlag}, \bibinfo{address}{Berlin}.
\newblock
\showISBNx{9783540136279}
\showLCCN{88127075}
\urldef\tempurl%
\url{https://books.google.com/books?id=S4wnAQAAIAAJ}
\showURL{%
\tempurl}


\bibitem[Chen et~al\mbox{.}(2022)]%
        {Chen_2022_BMVC}
\bibfield{author}{\bibinfo{person}{Meida Chen}, \bibinfo{person}{Qingyong Hu},
  \bibinfo{person}{Zifan Yu}, \bibinfo{person}{Hugues THOMAS},
  \bibinfo{person}{Andrew Feng}, \bibinfo{person}{Yu Hou},
  \bibinfo{person}{Kyle McCullough}, \bibinfo{person}{Fengbo Ren}, {and}
  \bibinfo{person}{Lucio Soibelman}.} \bibinfo{year}{2022}\natexlab{}.
\newblock \showarticletitle{STPLS3D: A Large-Scale Synthetic and Real Aerial
  Photogrammetry 3D Point Cloud Dataset}. In \bibinfo{booktitle}{\emph{33rd
  British Machine Vision Conference, November 21-24, 2022}}.
  \bibinfo{publisher}{{BMVA} Press}, \bibinfo{address}{London, UK},
  \bibinfo{pages}{429}.
\newblock
\urldef\tempurl%
\url{https://bmvc2022.mpi-inf.mpg.de/0429.pdf}
\showURL{%
\tempurl}


\bibitem[Cho et~al\mbox{.}(2023)]%
        {gowalla}
\bibfield{author}{\bibinfo{person}{E. Cho}, \bibinfo{person}{S.~A. Myers},
  {and} \bibinfo{person}{J. Leskoven}.} \bibinfo{year}{2023}\natexlab{}.
\newblock \bibinfo{title}{Friendship and Mobility: User Movement in
  Location-Based Social Networks}.
\newblock
\newblock
\newblock
\shownote{Retrieved from UCR-STAR
  \url{https://star.cs.ucr.edu/?stanford-gowalla&d}}.


\bibitem[Evangelou et~al\mbox{.}(2021)]%
        {evangelou}
\bibfield{author}{\bibinfo{person}{I. Evangelou}, \bibinfo{person}{G.
  Papaioannou}, \bibinfo{person}{K. Vardis}, {and} \bibinfo{person}{A.~A.
  Vasilakis}.} \bibinfo{year}{2021}\natexlab{}.
\newblock \showarticletitle{Fast Radius Search Exploiting Ray Tracing
  Frameworks}.
\newblock \bibinfo{journal}{\emph{Journal of Computer Graphics Techniques
  (JCGT)}} \bibinfo{volume}{10}, \bibinfo{number}{1} (\bibinfo{date}{5
  February} \bibinfo{year}{2021}), \bibinfo{pages}{25--48}.
\newblock
\showISSN{2331-7418}
\urldef\tempurl%
\url{http://jcgt.org/published/0010/01/02/}
\showURL{%
\tempurl}


\bibitem[Friedman et~al\mbox{.}(1977)]%
        {friedman1977algorithm}
\bibfield{author}{\bibinfo{person}{Jerome~H Friedman},
  \bibinfo{person}{Jon~Louis Bentley}, {and} \bibinfo{person}{Raphael~Ari
  Finkel}.} \bibinfo{year}{1977}\natexlab{}.
\newblock \showarticletitle{An algorithm for finding best matches in
  logarithmic expected time}.
\newblock \bibinfo{journal}{\emph{ACM Transactions on Mathematical Software
  (TOMS)}} \bibinfo{volume}{3}, \bibinfo{number}{3} (\bibinfo{year}{1977}),
  \bibinfo{pages}{209--226}.
\newblock


\bibitem[Fu et~al\mbox{.}(2019)]%
        {nsg}
\bibfield{author}{\bibinfo{person}{Cong Fu}, \bibinfo{person}{Chao Xiang},
  \bibinfo{person}{Changxu Wang}, {and} \bibinfo{person}{Deng Cai}.}
  \bibinfo{year}{2019}\natexlab{}.
\newblock \showarticletitle{Fast Approximate Nearest Neighbor Search With The
  Navigating Spreading-out Graph}.
\newblock \bibinfo{journal}{\emph{Proc. {VLDB} Endow.}} \bibinfo{volume}{12},
  \bibinfo{number}{5} (\bibinfo{year}{2019}), \bibinfo{pages}{461--474}.
\newblock


\bibitem[{GBIF.Org User}(2023)]%
        {gbif}
\bibfield{author}{\bibinfo{person}{{GBIF.Org User}}.}
  \bibinfo{year}{2023}\natexlab{}.
\newblock \bibinfo{title}{Occurrence Download}.
\newblock
\newblock
\urldef\tempurl%
\url{https://doi.org/10.15468/DL.QQ7KRQ}
\showDOI{\tempurl}


\bibitem[Geiger et~al\mbox{.}(2013)]%
        {kitti}
\bibfield{author}{\bibinfo{person}{Andreas Geiger}, \bibinfo{person}{Philip
  Lenz}, \bibinfo{person}{Christoph Stiller}, {and} \bibinfo{person}{Raquel
  Urtasun}.} \bibinfo{year}{2013}\natexlab{}.
\newblock \bibinfo{title}{Vision meets Robotics: The KITTI Dataset}.
\newblock
\newblock
\urldef\tempurl%
\url{https://www.cvlibs.net/datasets/kitti/raw_data.php}
\showURL{%
\tempurl}


\bibitem[Gieseke et~al\mbox{.}(2014)]%
        {bufferkdtree}
\bibfield{author}{\bibinfo{person}{Fabian Gieseke}, \bibinfo{person}{Justin
  Heinermann}, \bibinfo{person}{Cosmin~E. Oancea}, {and}
  \bibinfo{person}{Christian Igel}.} \bibinfo{year}{2014}\natexlab{}.
\newblock \showarticletitle{Buffer k-d Trees: Processing Massive Nearest
  Neighbor Queries on GPUs}. In \bibinfo{booktitle}{\emph{{ICML}}}
  \emph{(\bibinfo{series}{{JMLR} Workshop and Conference Proceedings},
  Vol.~\bibinfo{volume}{32})}. \bibinfo{publisher}{JMLR.org},
  \bibinfo{pages}{172--180}.
\newblock


\bibitem[Goldfarb et~al\mbox{.}(2013)]%
        {autoropes}
\bibfield{author}{\bibinfo{person}{Michael Goldfarb},
  \bibinfo{person}{Youngjoon Jo}, {and} \bibinfo{person}{Milind Kulkarni}.}
  \bibinfo{year}{2013}\natexlab{}.
\newblock \showarticletitle{General Transformations for GPU Execution of Tree
  Traversals}. In \bibinfo{booktitle}{\emph{Proceedings of the International
  Conference on High Performance Computing, Networking, Storage and Analysis}}
  (Denver, Colorado) \emph{(\bibinfo{series}{SC '13})}.
  \bibinfo{publisher}{Association for Computing Machinery},
  \bibinfo{address}{New York, NY, USA}, Article \bibinfo{articleno}{10},
  \bibinfo{numpages}{12}~pages.
\newblock
\showISBNx{9781450323789}
\urldef\tempurl%
\url{https://doi.org/10.1145/2503210.2503223}
\showDOI{\tempurl}


\bibitem[Gowanlock(2021)]%
        {gowanlock2021hybrid}
\bibfield{author}{\bibinfo{person}{Michael Gowanlock}.}
  \bibinfo{year}{2021}\natexlab{}.
\newblock \showarticletitle{Hybrid KNN-join: Parallel nearest neighbor searches
  exploiting CPU and GPU architectural features}.
\newblock \bibinfo{journal}{\emph{J. Parallel and Distrib. Comput.}}
  \bibinfo{volume}{149} (\bibinfo{year}{2021}), \bibinfo{pages}{119--137}.
\newblock


\bibitem[Guo et~al\mbox{.}(2019)]%
        {scann}
\bibfield{author}{\bibinfo{person}{Ruiqi Guo}, \bibinfo{person}{Philip Sun},
  \bibinfo{person}{Erik Lindgren}, \bibinfo{person}{Quan Geng},
  \bibinfo{person}{David Simcha}, \bibinfo{person}{Felix Chern}, {and}
  \bibinfo{person}{Sanjiv Kumar}.} \bibinfo{year}{2019}\natexlab{}.
\newblock \bibinfo{title}{Accelerating Large-Scale Inference with Anisotropic
  Vector Quantization}.
\newblock
\newblock
\urldef\tempurl%
\url{https://doi.org/10.48550/ARXIV.1908.10396}
\showDOI{\tempurl}


\bibitem[Hamming(1950)]%
        {hamming}
\bibfield{author}{\bibinfo{person}{R.~W. Hamming}.}
  \bibinfo{year}{1950}\natexlab{}.
\newblock \showarticletitle{Error detecting and error correcting codes}.
\newblock \bibinfo{journal}{\emph{The Bell System Technical Journal}}
  \bibinfo{volume}{29}, \bibinfo{number}{2} (\bibinfo{year}{1950}),
  \bibinfo{pages}{147--160}.
\newblock
\urldef\tempurl%
\url{https://doi.org/10.1002/j.1538-7305.1950.tb00463.x}
\showDOI{\tempurl}


\bibitem[Hegde et~al\mbox{.}(2017)]%
        {treelogy}
\bibfield{author}{\bibinfo{person}{Nikhil Hegde}, \bibinfo{person}{Jianqiao
  Liu}, \bibinfo{person}{Kirshanthan Sundararajah}, {and}
  \bibinfo{person}{Milind Kulkarni}.} \bibinfo{year}{2017}\natexlab{}.
\newblock \showarticletitle{Treelogy: A benchmark suite for tree traversals}.
  In \bibinfo{booktitle}{\emph{2017 IEEE International Symposium on Performance
  Analysis of Systems and Software (ISPASS)}}. \bibinfo{pages}{227--238}.
\newblock
\urldef\tempurl%
\url{https://doi.org/10.1109/ISPASS.2017.7975294}
\showDOI{\tempurl}


\bibitem[Hu et~al\mbox{.}(2020)]%
        {hu2019randla}
\bibfield{author}{\bibinfo{person}{Qingyong Hu}, \bibinfo{person}{Bo Yang},
  \bibinfo{person}{Linhai Xie}, \bibinfo{person}{Stefano Rosa},
  \bibinfo{person}{Yulan Guo}, \bibinfo{person}{Zhihua Wang},
  \bibinfo{person}{Niki Trigoni}, {and} \bibinfo{person}{Andrew Markham}.}
  \bibinfo{year}{2020}\natexlab{}.
\newblock \showarticletitle{RandLA-Net: Efficient Semantic Segmentation of
  Large-Scale Point Clouds}.
\newblock \bibinfo{journal}{\emph{Proceedings of the IEEE Conference on
  Computer Vision and Pattern Recognition}} (\bibinfo{year}{2020}).
\newblock


\bibitem[Huang et~al\mbox{.}(2015)]%
        {huang2015query}
\bibfield{author}{\bibinfo{person}{Qiang Huang}, \bibinfo{person}{Jianlin
  Feng}, \bibinfo{person}{Yikai Zhang}, \bibinfo{person}{Qiong Fang}, {and}
  \bibinfo{person}{Wilfred Ng}.} \bibinfo{year}{2015}\natexlab{}.
\newblock \showarticletitle{Query-aware locality-sensitive hashing for
  approximate nearest neighbor search}.
\newblock \bibinfo{journal}{\emph{Proceedings of the VLDB Endowment}}
  \bibinfo{volume}{9}, \bibinfo{number}{1} (\bibinfo{year}{2015}),
  \bibinfo{pages}{1--12}.
\newblock


\bibitem[Indyk and Motwani(1998)]%
        {indyk1998lsh}
\bibfield{author}{\bibinfo{person}{Piotr Indyk} {and} \bibinfo{person}{Rajeev
  Motwani}.} \bibinfo{year}{1998}\natexlab{}.
\newblock \showarticletitle{Approximate nearest neighbors: towards removing the
  curse of dimensionality}. In \bibinfo{booktitle}{\emph{Proceedings of the
  thirtieth annual ACM symposium on Theory of computing}}.
  \bibinfo{pages}{604--613}.
\newblock


\bibitem[Intel(2023)]%
        {intel}
\bibfield{author}{\bibinfo{person}{Intel}.} \bibinfo{year}{2023}\natexlab{}.
\newblock \bibinfo{title}{Intel Ray tracing}.
\newblock
\newblock
\urldef\tempurl%
\url{https://www.intel.com/content/www/us/en/developer/articles/guide/real-time-ray-tracing-in-games.html}
\showURL{%
\tempurl}


\bibitem[Jaccard(1912)]%
        {jaccard}
\bibfield{author}{\bibinfo{person}{Paul Jaccard}.}
  \bibinfo{year}{1912}\natexlab{}.
\newblock \showarticletitle{THE DISTRIBUTION OF THE FLORA IN THE ALPINE
  ZONE.1}.
\newblock \bibinfo{journal}{\emph{New Phytologist}} \bibinfo{volume}{11},
  \bibinfo{number}{2} (\bibinfo{year}{1912}), \bibinfo{pages}{37--50}.
\newblock
\urldef\tempurl%
\url{https://doi.org/10.1111/j.1469-8137.1912.tb05611.x}
\showDOI{\tempurl}


\bibitem[Johnson et~al\mbox{.}(2021)]%
        {faiss}
\bibfield{author}{\bibinfo{person}{J. Johnson}, \bibinfo{person}{M. Douze},
  {and} \bibinfo{person}{H. Jegou}.} \bibinfo{year}{2021}\natexlab{}.
\newblock \showarticletitle{Billion-Scale Similarity Search with GPUs}.
\newblock \bibinfo{journal}{\emph{IEEE Transactions on Big Data}}
  \bibinfo{volume}{7}, \bibinfo{number}{03} (\bibinfo{date}{Jul}
  \bibinfo{year}{2021}), \bibinfo{pages}{535--547}.
\newblock
\showISSN{2332-7790}
\urldef\tempurl%
\url{https://doi.org/10.1109/TBDATA.2019.2921572}
\showDOI{\tempurl}


\bibitem[Kaiser and Sutskever(2015)]%
        {neural}
\bibfield{author}{\bibinfo{person}{Lukasz Kaiser} {and} \bibinfo{person}{Ilya
  Sutskever}.} \bibinfo{year}{2015}\natexlab{}.
\newblock \bibinfo{title}{Neural GPUs Learn Algorithms}.
\newblock
\newblock
\urldef\tempurl%
\url{https://doi.org/10.48550/ARXIV.1511.08228}
\showDOI{\tempurl}


\bibitem[Mahalanobis(1936)]%
        {mahalanobis}
\bibfield{author}{\bibinfo{person}{Prasanta~Chandra Mahalanobis}.}
  \bibinfo{year}{1936}\natexlab{}.
\newblock \bibinfo{title}{On the generalised distance in statistics.}
\newblock
\newblock
\urldef\tempurl%
\url{http://library.isical.ac.in:8080/xmlui/bitstream/handle/10263/6765/Vol02_1936_1_Art05-pcm.pdf}
\showURL{%
\tempurl}


\bibitem[Malkov and Yashunin(2020)]%
        {hnsw}
\bibfield{author}{\bibinfo{person}{Yu~A. Malkov} {and} \bibinfo{person}{D.~A.
  Yashunin}.} \bibinfo{year}{2020}\natexlab{}.
\newblock \showarticletitle{Efficient and Robust Approximate Nearest Neighbor
  Search Using Hierarchical Navigable Small World Graphs}.
\newblock \bibinfo{journal}{\emph{IEEE Trans. Pattern Anal. Mach. Intell.}}
  \bibinfo{volume}{42}, \bibinfo{number}{4} (\bibinfo{date}{apr}
  \bibinfo{year}{2020}), \bibinfo{pages}{824–836}.
\newblock
\showISSN{0162-8828}
\urldef\tempurl%
\url{https://doi.org/10.1109/TPAMI.2018.2889473}
\showDOI{\tempurl}


\bibitem[Merry et~al\mbox{.}(2013)]%
        {kdtree-knn}
\bibfield{author}{\bibinfo{person}{Bruce Merry}, \bibinfo{person}{James Gain},
  {and} \bibinfo{person}{Patrick Marais}.} \bibinfo{year}{2013}\natexlab{}.
\newblock \showarticletitle{{Accelerating kd-tree Searches for all k-nearest
  Neighbours}}. In \bibinfo{booktitle}{\emph{Eurographics 2013 - Short
  Papers}}, \bibfield{editor}{\bibinfo{person}{M.-A. Otaduy} {and}
  \bibinfo{person}{O.~Sorkine}} (Eds.). \bibinfo{publisher}{The Eurographics
  Association}.
\newblock
\showISSN{1017-4656}
\urldef\tempurl%
\url{https://doi.org/10.2312/conf/EG2013/short/037-040}
\showDOI{\tempurl}


\bibitem[Mori et~al\mbox{.}(2001)]%
        {vision}
\bibfield{author}{\bibinfo{person}{G. Mori}, \bibinfo{person}{S. Belongie},
  {and} \bibinfo{person}{J. Malik}.} \bibinfo{year}{2001}\natexlab{}.
\newblock \showarticletitle{Shape contexts enable efficient retrieval of
  similar shapes}. In \bibinfo{booktitle}{\emph{Proceedings of the 2001 IEEE
  Computer Society Conference on Computer Vision and Pattern Recognition. CVPR
  2001}}, Vol.~\bibinfo{volume}{1}. \bibinfo{pages}{I--I}.
\newblock
\urldef\tempurl%
\url{https://doi.org/10.1109/CVPR.2001.990547}
\showDOI{\tempurl}


\bibitem[Muja and Lowe(2014)]%
        {flann}
\bibfield{author}{\bibinfo{person}{Marius Muja} {and} \bibinfo{person}{David~G.
  Lowe}.} \bibinfo{year}{2014}\natexlab{}.
\newblock \showarticletitle{Scalable Nearest Neighbor Algorithms for High
  Dimensional Data}.
\newblock \bibinfo{journal}{\emph{IEEE Transactions on Pattern Analysis and
  Machine Intelligence}} \bibinfo{volume}{36}, \bibinfo{number}{11}
  (\bibinfo{year}{2014}), \bibinfo{pages}{2227--2240}.
\newblock
\urldef\tempurl%
\url{https://doi.org/10.1109/TPAMI.2014.2321376}
\showDOI{\tempurl}


\bibitem[Nagarajan and Kulkarni(2023)]%
        {dbscan}
\bibfield{author}{\bibinfo{person}{Vani Nagarajan} {and}
  \bibinfo{person}{Milind Kulkarni}.} \bibinfo{year}{2023}\natexlab{}.
\newblock \showarticletitle{{RT-DBSCAN:} Accelerating {DBSCAN} using Ray
  Tracing Hardware}. In \bibinfo{booktitle}{\emph{{IPDPS}}}.
  \bibinfo{publisher}{{IEEE}}, \bibinfo{pages}{963--973}.
\newblock


\bibitem[Nagarajan et~al\mbox{.}(2023)]%
        {trueknn}
\bibfield{author}{\bibinfo{person}{Vani Nagarajan}, \bibinfo{person}{Durga
  Mandarapu}, {and} \bibinfo{person}{Milind Kulkarni}.}
  \bibinfo{year}{2023}\natexlab{}.
\newblock \showarticletitle{RT-kNNS Unbound: Using RT Cores to Accelerate
  Unrestricted Neighbor Search}. In \bibinfo{booktitle}{\emph{Proceedings of
  the 37th International Conference on Supercomputing, {ICS} 2023, Orlando, FL,
  USA, June 21-23, 2023}}, \bibfield{editor}{\bibinfo{person}{Kyle~A.
  Gallivan}, \bibinfo{person}{Efstratios Gallopoulos},
  \bibinfo{person}{Dimitrios~S. Nikolopoulos}, {and}
  \bibinfo{person}{Ram{\'{o}}n Beivide}} (Eds.). \bibinfo{publisher}{{ACM}},
  \bibinfo{pages}{289--300}.
\newblock
\urldef\tempurl%
\url{https://doi.org/10.1145/3577193.3593738}
\showDOI{\tempurl}


\bibitem[Nam et~al\mbox{.}(2016)]%
        {nam2016parallel}
\bibfield{author}{\bibinfo{person}{Moohyeon Nam}, \bibinfo{person}{Jinwoong
  Kim}, {and} \bibinfo{person}{Beomseok Nam}.} \bibinfo{year}{2016}\natexlab{}.
\newblock \showarticletitle{Parallel tree traversal for nearest neighbor query
  on the GPU}. In \bibinfo{booktitle}{\emph{2016 45th International Conference
  on Parallel Processing (ICPP)}}. IEEE, \bibinfo{pages}{113--122}.
\newblock


\bibitem[NVIDIA({[n.\,d.]})]%
        {optix}
\bibfield{author}{\bibinfo{person}{NVIDIA}.}
  \bibinfo{year}{[n.\,d.]}\natexlab{}.
\newblock \bibinfo{title}{NVIDIA OptiX 7.5 – Programming Guide}.
\newblock
\newblock
\urldef\tempurl%
\url{https://raytracing-docs.nvidia.com/optix7/guide/index.html}
\showURL{%
\tempurl}


\bibitem[Nvidia(2023a)]%
        {nsight}
\bibfield{author}{\bibinfo{person}{Nvidia}.} \bibinfo{year}{2023}\natexlab{a}.
\newblock \bibinfo{title}{NVIDIA Nsight Compute}.
\newblock
\newblock
\urldef\tempurl%
\url{https://developer.nvidia.com/nsight-compute}
\showURL{%
\tempurl}


\bibitem[Nvidia(2023b)]%
        {nvidia}
\bibfield{author}{\bibinfo{person}{Nvidia}.} \bibinfo{year}{2023}\natexlab{b}.
\newblock \bibinfo{title}{NVIDIA Ray tracing}.
\newblock
\newblock
\urldef\tempurl%
\url{https://developer.nvidia.com/rtx/ray-tracing}
\showURL{%
\tempurl}


\bibitem[OpenStreetMap({[n.\,d.]})]%
        {osm}
\bibfield{author}{\bibinfo{person}{OpenStreetMap}.}
  \bibinfo{year}{[n.\,d.]}\natexlab{}.
\newblock
\newblock
\urldef\tempurl%
\url{https://www.openstreetmap.org}
\showURL{%
\tempurl}


\bibitem[Pandey et~al\mbox{.}(2018)]%
        {pandey2018good}
\bibfield{author}{\bibinfo{person}{Varun Pandey}, \bibinfo{person}{Andreas
  Kipf}, \bibinfo{person}{Thomas Neumann}, {and} \bibinfo{person}{Alfons
  Kemper}.} \bibinfo{year}{2018}\natexlab{}.
\newblock \showarticletitle{How good are modern spatial analytics systems?}
\newblock \bibinfo{journal}{\emph{Proceedings of the VLDB Endowment}}
  \bibinfo{volume}{11}, \bibinfo{number}{11} (\bibinfo{year}{2018}),
  \bibinfo{pages}{1661--1673}.
\newblock


\bibitem[Pennington et~al\mbox{.}(2014)]%
        {glove}
\bibfield{author}{\bibinfo{person}{Jeffrey Pennington},
  \bibinfo{person}{Richard Socher}, {and} \bibinfo{person}{Christopher~D.
  Manning}.} \bibinfo{year}{2014}\natexlab{}.
\newblock \showarticletitle{GloVe: Global Vectors for Word Representation}. In
  \bibinfo{booktitle}{\emph{Empirical Methods in Natural Language Processing
  (EMNLP)}}. \bibinfo{pages}{1532--1543}.
\newblock
\urldef\tempurl%
\url{http://www.aclweb.org/anthology/D14-1162}
\showURL{%
\tempurl}


\bibitem[Pham and Liu(2022)]%
        {pham2022falconn}
\bibfield{author}{\bibinfo{person}{Ninh Pham} {and} \bibinfo{person}{Tao Liu}.}
  \bibinfo{year}{2022}\natexlab{}.
\newblock \bibinfo{title}{Falconn++: A Locality-sensitive Filtering Approach
  for Approximate Nearest Neighbor Search}.
\newblock
\newblock
\showeprint[arxiv]{2206.01382}~[cs.DS]


\bibitem[Qiu et~al\mbox{.}(2009)]%
        {qiu2009gpu}
\bibfield{author}{\bibinfo{person}{Deyuan Qiu}, \bibinfo{person}{Stefan May},
  {and} \bibinfo{person}{Andreas N{\"u}chter}.}
  \bibinfo{year}{2009}\natexlab{}.
\newblock \showarticletitle{GPU-accelerated nearest neighbor search for 3D
  registration}. In \bibinfo{booktitle}{\emph{Computer Vision Systems: 7th
  International Conference on Computer Vision Systems, ICVS 2009 Li{\`e}ge,
  Belgium, October 13-15, 2009. Proceedings 7}}. Springer,
  \bibinfo{pages}{194--203}.
\newblock


\bibitem[Reid and Menten(2020)]%
        {parallax}
\bibfield{author}{\bibinfo{person}{Mark~J. Reid} {and} \bibinfo{person}{Karl~M.
  Menten}.} \bibinfo{year}{2020}\natexlab{}.
\newblock \showarticletitle{The first stellar parallaxes revisited}.
\newblock \bibinfo{journal}{\emph{Astronomische Nachrichten}}
  \bibinfo{volume}{341}, \bibinfo{number}{9} (\bibinfo{date}{nov}
  \bibinfo{year}{2020}), \bibinfo{pages}{860--869}.
\newblock
\urldef\tempurl%
\url{https://doi.org/10.1002/asna.202013833}
\showDOI{\tempurl}


\bibitem[Repository(2014)]%
        {manuscript}
\bibfield{author}{\bibinfo{person}{The Stanford 3D~Scanning Repository}.}
  \bibinfo{year}{2014}\natexlab{}.
\newblock \bibinfo{title}{Vellum manuscript, The XYZ RGB models}.
\newblock
\newblock
\urldef\tempurl%
\url{http://graphics.stanford.edu/data/3Dscanrep/}
\showURL{%
\tempurl}


\bibitem[Rubin and Whitted(1980)]%
        {bvh_first}
\bibfield{author}{\bibinfo{person}{Steven Rubin} {and} \bibinfo{person}{Turner
  Whitted}.} \bibinfo{year}{1980}\natexlab{}.
\newblock \showarticletitle{A 3-dimensional representation for fast rendering
  of complex scenes}.
\newblock \bibinfo{journal}{\emph{ACM Siggraph Computer Graphics}}
  \bibinfo{volume}{14}.
\newblock
\urldef\tempurl%
\url{https://doi.org/10.1145/965105.807479}
\showDOI{\tempurl}


\bibitem[Singhal(2001)]%
        {cosine}
\bibfield{author}{\bibinfo{person}{Amit Singhal}.}
  \bibinfo{year}{2001}\natexlab{}.
\newblock \showarticletitle{Modern Information Retrieval: {A} Brief Overview}.
\newblock \bibinfo{journal}{\emph{{IEEE} Data Eng. Bull.}}
  \bibinfo{volume}{24}, \bibinfo{number}{4} (\bibinfo{year}{2001}),
  \bibinfo{pages}{35--43}.
\newblock
\urldef\tempurl%
\url{http://sites.computer.org/debull/A01DEC-CD.pdf}
\showURL{%
\tempurl}


\bibitem[Wald et~al\mbox{.}({[n.\,d.]})]%
        {owl}
\bibfield{author}{\bibinfo{person}{Ingo Wald}, \bibinfo{person}{Nathan
  Morrical}, {and} \bibinfo{person}{Haines E}.}
  \bibinfo{year}{[n.\,d.]}\natexlab{}.
\newblock \bibinfo{title}{OWL: A Node Graph "Wrapper" Library for OptiX 7}.
\newblock
\newblock
\urldef\tempurl%
\url{https://github.com/owl-project/owl}
\showURL{%
\tempurl}


\bibitem[Wald et~al\mbox{.}(2019)]%
        {wald}
\bibfield{author}{\bibinfo{person}{Ingo Wald}, \bibinfo{person}{Will Usher},
  \bibinfo{person}{Nathan Morrical}, \bibinfo{person}{Laura Lediaev}, {and}
  \bibinfo{person}{Valerio Pascucci}.} \bibinfo{year}{2019}\natexlab{}.
\newblock \showarticletitle{{RTX Beyond Ray Tracing: Exploring the Use of
  Hardware Ray Tracing Cores for Tet-Mesh Point Location}}. In
  \bibinfo{booktitle}{\emph{High-Performance Graphics - Short Papers}},
  \bibfield{editor}{\bibinfo{person}{Markus Steinberger} {and}
  \bibinfo{person}{Tim Foley}} (Eds.). \bibinfo{publisher}{The Eurographics
  Association}.
\newblock
\showISBNx{978-3-03868-092-5}
\showISSN{2079-8687}
\urldef\tempurl%
\url{https://doi.org/10.2312/hpg.20191189}
\showDOI{\tempurl}


\bibitem[Weber et~al\mbox{.}(1998)]%
        {vafile}
\bibfield{author}{\bibinfo{person}{Roger Weber}, \bibinfo{person}{Hans-J{\"o}rg
  Schek}, {and} \bibinfo{person}{Stephen Blott}.}
  \bibinfo{year}{1998}\natexlab{}.
\newblock \showarticletitle{A Quantitative Analysis and Performance Study for
  Similarity-Search Methods in High-Dimensional Spaces}. In
  \bibinfo{booktitle}{\emph{VLDB}}.
\newblock


\bibitem[Wood(2008)]%
        {filter}
\bibfield{author}{\bibinfo{person}{Jordan Wood}.}
  \bibinfo{year}{2008}\natexlab{}.
\newblock \bibinfo{booktitle}{\emph{Filter and Refine Strategy}}.
\newblock \bibinfo{publisher}{Springer US}, \bibinfo{address}{Boston, MA},
  \bibinfo{pages}{320--320}.
\newblock
\showISBNx{978-0-387-35973-1}
\urldef\tempurl%
\url{https://doi.org/10.1007/978-0-387-35973-1_415}
\showDOI{\tempurl}


\bibitem[Xu et~al\mbox{.}(2019)]%
        {tigris}
\bibfield{author}{\bibinfo{person}{Tiancheng Xu}, \bibinfo{person}{Boyuan
  Tian}, {and} \bibinfo{person}{Yuhao Zhu}.} \bibinfo{year}{2019}\natexlab{}.
\newblock \showarticletitle{Tigris: Architecture and Algorithms for 3D
  Perception in Point Clouds}. In \bibinfo{booktitle}{\emph{Proceedings of the
  52nd Annual IEEE/ACM International Symposium on Microarchitecture}}
  (Columbus, OH, USA) \emph{(\bibinfo{series}{MICRO '52})}.
  \bibinfo{publisher}{Association for Computing Machinery},
  \bibinfo{address}{New York, NY, USA}, \bibinfo{pages}{629–642}.
\newblock
\showISBNx{9781450369381}
\urldef\tempurl%
\url{https://doi.org/10.1145/3352460.3358259}
\showDOI{\tempurl}


\bibitem[Zellmann et~al\mbox{.}(2020)]%
        {zellman}
\bibfield{author}{\bibinfo{person}{Stefan Zellmann}, \bibinfo{person}{Martin
  Weier}, {and} \bibinfo{person}{Ingo Wald}.} \bibinfo{year}{2020}\natexlab{}.
\newblock \showarticletitle{Accelerating Force-Directed Graph Drawing with RT
  Cores}. In \bibinfo{booktitle}{\emph{2020 IEEE Visualization Conference
  (VIS)}}. \bibinfo{pages}{96--100}.
\newblock
\urldef\tempurl%
\url{https://doi.org/10.1109/VIS47514.2020.00026}
\showDOI{\tempurl}


\bibitem[Zhao et~al\mbox{.}(2020)]%
        {song}
\bibfield{author}{\bibinfo{person}{Weijie Zhao}, \bibinfo{person}{Shulong Tan},
  {and} \bibinfo{person}{Ping Li}.} \bibinfo{year}{2020}\natexlab{}.
\newblock \showarticletitle{SONG: Approximate Nearest Neighbor Search on GPU}.
  In \bibinfo{booktitle}{\emph{2020 IEEE 36th International Conference on Data
  Engineering (ICDE)}}. \bibinfo{pages}{1033--1044}.
\newblock
\urldef\tempurl%
\url{https://doi.org/10.1109/ICDE48307.2020.00094}
\showDOI{\tempurl}


\bibitem[Zhu(2022)]%
        {rtnn}
\bibfield{author}{\bibinfo{person}{Yuhao Zhu}.}
  \bibinfo{year}{2022}\natexlab{}.
\newblock \showarticletitle{RTNN: Accelerating Neighbor Search Using Hardware
  Ray Tracing}. In \bibinfo{booktitle}{\emph{Proceedings of the 27th ACM
  SIGPLAN Symposium on Principles and Practice of Parallel Programming}}
  (Seoul, Republic of Korea) \emph{(\bibinfo{series}{PPoPP '22})}.
  \bibinfo{publisher}{Association for Computing Machinery},
  \bibinfo{address}{New York, NY, USA}, \bibinfo{pages}{76–89}.
\newblock
\showISBNx{9781450392044}
\urldef\tempurl%
\url{https://doi.org/10.1145/3503221.3508409}
\showDOI{\tempurl}


\end{thebibliography}

\end{document}